\documentclass[aps,prd,reprint,longbibliography,nofootinbib,superscriptaddress,floatfix]{revtex4-2}

\usepackage{subcaption}
\usepackage{comment}
\usepackage{slashed}
\usepackage{verbatim}
\usepackage[T1]{fontenc}
\usepackage{comment}
\usepackage{mathbbol}
\usepackage{orcidlink}
\usepackage[normalem]{ulem}
\usepackage[english]{babel}
\usepackage{physics}
\usepackage{subcaption}
\usepackage{dcolumn}
\usepackage{tensor}
\usepackage{comment}
\usepackage{placeins}
\usepackage{graphicx,overpic,mathtools}
\usepackage{amsthm,amsmath,amssymb,mathrsfs}
\usepackage{braket,bm,bbm,setspace}
\usepackage{booktabs}
\usepackage{cancel}
\usepackage{float}
\usepackage{xargs}
\usepackage{ragged2e}

\usepackage[dvipsnames,table,xcdraw]{xcolor}

\usepackage{hyperref}
\hypersetup{
    colorlinks=true,
    linkcolor=RoyalBlue,
    citecolor=ForestGreen,
    urlcolor=RoyalBlue}

\usepackage[nameinlink,capitalise]{cleveref}

\newcommand{\uib}{%
Departament de F\'isica, Universitat de les Illes Balears,
IAC3~--~IEEC, Crta.\ Valldemossa km~7.5, E-07122 Palma, Spain}
\newcommand{\GSSI}{%
Gran Sasso Science Institute (GSSI), I-67100 L'Aquila, Italy}
\newcommand{\GranSasso}{%
INFN, Laboratori Nazionali del Gran Sasso, I-67100 Assergi, Italy}

\begin{document}
%%%%%%%%%%%%%%%%%%%%%%%%%%%%%%%%%%%%%%%%%%%%%%%%%%%%%%%%%%
\title{Shaping black hole resonances II: \\ Spectroscopic landscapes and interference in black hole ringdown}

\author{Alejandro Svyatkovskyy Kholyavka\orcidlink{0009-0004-1688-6018}}
\email{alejandro.svyatkovskyy@uib.cat}
\affiliation{\uib}

\author{Sayak Datta\orcidlink{0000-0002-4774-0298}}
\email{sayak.datta@gssi.it}
\affiliation{\GSSI}
\affiliation{\GranSasso}

\author{Jose Antonio Le\'on Vega\orcidlink{0009-0002-7382-5137}}
\email{jose-antonio.leon1@estudiant.uib.cat}
\affiliation{\uib}

\author{Xisco Jim\'enez Forteza\orcidlink{0000-0002-8158-5009}}
\email{f.jimenez@uib.es}
\affiliation{\uib}

%%%%%%%%%%%%%%%%%%%%%%%%%%%%%%%%%%%%%%%%%%%%%%%%%%%%%%%%%%

\begin{abstract}
We map analytically the excitation of Schwarzschild quasinormal modes (QNMs) $C_n$ by oscillatory Gaussian perturbations across the pulse width $\sigma$ and carrier frequency $\nu$. The excitation coefficient of each mode is determined by the product of an intrinsic excitation factor and a source-dependent overlap evaluated at the complex QNM frequency. For oscillatory Gaussian perturbations, this overlap is the coherent sum of the positive- and negative-frequency branches of the source, with its amplitude factorizing exactly into a Gaussian spectral envelope and an interference factor. We observe that the excitation amplitude peaks at $\nu=\omega_n^{\rm Re}$ only for narrowband sources. 
For sufficiently broadband pulses, the maximum lies at $\nu=0$, so efficient excitation does not necessarily require matching the carrier and the mode frequency. We observe that the interference between branches, modulates the amplitude only for nonzero $\nu$, a feature generic to real oscillatory sources. The prominence of the interference-induced modulation is controlled by the ratio $\omega_n^{\rm Re}/\omega_n^{\rm Im}$. The phase separates into a mode-only, a width-only and a $\nu$-dependent term, the last generated by the same interference, with $\nu=0$ being its unique maximum at fixed width. The analytical predictions are validated against time-domain numerical simulations, showing a relative median error of $\sim 4\%$ for $|C_0|$. For $|C_1|$, the median error is $\sim 16\%$ when $\sigma\leq 5$, increasing to $\sim 36\%$ across the full parameter space. The median phase difference for $n=0$  is $\sim0.14 \,\rm rad $ and for $n=1$, $\sigma\leq5$ it is $0.34\,\rm rad$. Our results demonstrate that QNM excitation cannot be understood through frequency matching alone and may provide qualitative insights into more realistic astrophysical scenarios.
\end{abstract}

\maketitle
\setlength{\parskip}{0.5em}

%%%%%%%%%%%%%%%%%%%%%%%%%%%%%%%%%%%%%%%%%%%%%%%%%%%%%%%%%%
%%%%%%%%%%%%%%%%%%%%%%%%%%%%%%%%%%%%%%%%%%%%%%%%%%%%%%%%%%
\section{Introduction}
\label{sec:intro}
%%%%%%%%%%%%%%%%%%%%%%%%%%%%%%%%%%%%%%%%%%%%%%%%%%%%%%%%%%
%%%%%%%%%%%%%%%%%%%%%%%%%%%%%%%%%%%%%%%%%%%%%%%%%%%%%%%%%%

Black hole (BH) ringdown offers one of the cleanest windows into the strong-field regime of gravity. Following a perturbation, a BH relaxes toward equilibrium by radiating a superposition of damped sinusoids, its quasinormal modes (QNMs)~\cite{vishveshwara:1970zz, press:1971wr, chandrasekhar:1975zza, kokkotas:1999bd, berti:2009kk, Konoplya:2011qq}. Since the frequencies of these modes are fixed entirely by the mass and spin of the final BH~\cite{detweiler:1980gk, Dreyer_2004, berti:2005ys}, comparing the observed ringdown spectrum against the QNM spectrum predicted by general relativity provides a direct, model-independent test of the theory~\cite{isi2019, Carullo:2019flq, Abbott:2020jks, Berti:2025hly}. As gravitational-wave detectors resolve increasingly subtle features~\cite{LIGOScientific:2025rid, LIGOScientific:2025wao, ET:2025xjr, Evans:2021gyd} of individual ringdowns, it becomes possible in principle to extract not only the mode frequencies $\omega_n$ but also how strongly each mode is excited~\cite{kamaretsos:2011um, london:2014cma, giesler2019, forteza2020, Cheung:2023vki}, information that is currently underused.

Within BH perturbation theory, QNM excitation is characterized by two complementary contributions: the intrinsic excitation factors, determined by the background spacetime, and the overlap between the perturbation and the QNM wavefunctions~\cite{Berti:2006wq,Zhang:2013ksa,Oshita:2021iyn,Lo:2025njp}. Recent studies have revealed resonant enhancements of the intrinsic excitation factors associated with avoided crossings and exceptional points in the QNM spectrum, both for Kerr BHs and in systems with coupled degrees of freedom~\cite{Kubota:2025hjk,Motohashi:2024fwt,Takahashi:2025uwo}. Other studies have investigated the reconstruction of ringdown waveforms from QNM excitation factors~\cite{Oshita:2024wgt}, as well as the impact of spectral instabilities and destructive interference among QNMs on the resulting ringdown signal~\cite{Oshita:2025ibu}. Complementarily, QNM amplitudes and phases have been extracted through fits to numerical-relativity waveforms, providing insights into their excitation during BH mergers and the challenges of disentangling individual modes~\cite{Finch:2021iip,Forteza:2021wfq,Forteza:2022tgq,Giesler:2024hcr,Mitman:2025hgy}. 

The overlap integrals between the perturbation and the mode structure provide a quantitative measure of how efficiently a given source excites individual QNMs~\cite{Leaver:1986gd, SunPrice1988, Andersson:1995zk, Berti:2006wq, Zhang:2013ksa, Oshita:2021iyn, Lo:2025njp, DellaRocca:2025zbe, Kubota:2025hjk}. Although appropriate within linear perturbation theory, this formulation has not been exploited sufficiently to reveal how excitation depends globally on the spectral properties of the source and thereby how to use it to infer the source properties.
In Ref.~\cite{Kholyavka:2026uam}, hereafter Paper~I, we showed that the overlap integral $T_n$ is the weighted spatial Fourier transform of the perturbation, evaluated at the complex QNM frequency $\omega_n$. A BH acts as a bank of resonant spectral filters. Depending on the source's spectral width, each filter selectively responds to the components of the source oscillating near its own frequency. Paper~I established this mechanism for a family of tunable, oscillatory Gaussian perturbations and validated it against time-domain numerical evolutions, while a similar formalism has been introduced to analyze the BH tails in~\cite{Vega:2026lfs} for Schwarzschild BHs and in~\cite{Gupta:2026szn} for Kerr.

In this work, we map QNM excitation across the source bandwidth $\sigma$ and carrier frequency $\nu$, revealing a structured landscape that differs between fundamental modes and overtones. We study the properties of the extrema of the mode amplitude response $C_n$ across the $\nu$ and $\sigma$ parameter space. We also explore the phase landscape of the excitation coefficient alongside the amplitude. Because the QNM frequency is complex, the contributions from the positive and negative branches carry a relative phase set by the damping rate, and their sum contains an interference term.  We further show that the phase has a distinct stationary structure. It contains unique extremum compared to the amplitude making it an independent observable. We derive these amplitude (Sec.~\ref{sec:Tn_structure}) and the phase landscape (Sec.~\ref{sec:phase}) properties analytically and test them against time-domain Regge--Wheeler (RW) evolutions for the fundamental mode and the first overtone. The amplitude maximum $|C_n|$ coincides with the QNM oscillation frequency only in the narrowband regime. For sufficiently broadband sources, the maximum can instead occur at zero carrier frequency. Moreover, the damping rate controls the stationary structure of the amplitude landscape, determining whether the QNMs should be excited maximally when the source frequency matches the QNM frequency.

The paper is organized as follows. In Sec.~\ref{sec:spectral_structure}, we introduce the spectral structure of the oscillatory source. In Sec.~\ref{sec:Tn_structure}, we analytically derive the excitation amplitude landscape, its stationary structure, and validate them with time-domain numerical simulations. In Sec.~\ref{sec:phase}, we extend the analysis to the phase of the excitation coefficient. We summarize our results and discuss their implications in Sec.~\ref{sec:conclusions}. We use $G=c=M=1$, write $\omega_n=\omega_n^{\rm Re}-i\,\omega_n^{\rm Im}$ with $\omega_n^{\rm Re},\omega_n^{\rm Im}>0$, and restrict to the $\ell=2$ Schwarzschild modes: $\omega_0=0.3737-0.0890i$, $\omega_1=0.3467-0.2739i$, $\omega_2=0.3011-0.4783i$ and $\omega_3=0.2515-0.7051i$~\cite{berti:2009kk}.

%%%%%%%%%%%%%%%%%%%%%%%%%%%%%%%%%%%%%%%%%%%%%%%%%%%%%%%%%%
%%%%%%%%%%%%%%%%%%%%%%%%%%%%%%%%%%%%%%%%%%%%%%%%%%%%%%%%%%
\section{Spectral structure of the source and its overlap}
\label{sec:spectral_structure}
%%%%%%%%%%%%%%%%%%%%%%%%%%%%%%%%%%%%%%%%%%%%%%%%%%%%%%%%%%
%%%%%%%%%%%%%%%%%%%%%%%%%%%%%%%%%%%%%%%%%%%%%%%%%%%%%%%%%%

%%%%%%%%%%%%%%%%%%%%%%%%%%%%%%%%%%%%%%%%%%%%%%%%%%%%%%%%%%
\subsection{Spectral filtration and initial data}
\label{sec:bifurcation}
%%%%%%%%%%%%%%%%%%%%%%%%%%%%%%%%%%%%%%%%%%%%%%%%%%%%%%%%%%

We work within the RW framework employed in Paper~I, to which we refer the reader for the complete Green's function (GF) formalism. At future null infinity, the QNM contribution to the waveform can be written as
\begin{equation}
\Psi\bigl(r_\star,t\bigr)=\sum_n C_n\,e^{-i\omega_n(t-r_\star)}\,,
\label{eq:ringdown}
\end{equation}
where $\omega_n$ denotes the complex frequency of the $n$-th QNM. The source-dependent contribution is given by the spectral overlap
\begin{equation}
T_n=\int_{-\infty}^{+\infty}\frac{\psi_{\rm in}\bigl(r'_\star,\omega_n\bigr)\,\mathcal{I}\bigl(\omega_n,r'_\star\bigr)}{A_{\rm out}\bigl(\omega_n\bigr)}\,\mathrm{d}r'_\star\,,
\label{eq:tn}
\end{equation}
which quantifies the overlap between the source term and the ingoing homogeneous solution evaluated at the QNM frequency. Here $\psi_{\rm in}$ is the solution of the homogeneous RW equation that is purely ingoing at the horizon, and $A_{\rm out}$ is the amplitude of its outgoing wave at infinity. The source term $\mathcal{I}$ is built from the initial data (ID).
When the overlap is dominated by the asymptotic region, Eq.~\eqref{eq:tn} reduces to
\begin{equation}
T_n=\int_{-\infty}^{+\infty}\mathcal{I}\bigl(\omega_n,r'_\star\bigr)\,e^{i\omega_n r'_\star}\,\mathrm{d}r'_\star\,.
\label{eq:tn_asymptotic}
\end{equation}
In this approximation, $T_n$ is the spatial Fourier transform of the perturbation evaluated at the complex QNM frequency. The quasinormal excitation coefficient (QNEC) can then be written as,
\begin{equation}
    C_n=B_n\,T_n\,
\end{equation}
where $B_n$ is the quasinormal excitation factor (QNEF), which depends on the BH and on the mode, not on the source parameters.

As demonstrated in Paper~I, the overlap is a filtered response of the source spectrum. In the asymptotic limit, it simplifies to $T_n = -2i\omega_n \Psi(\omega_n)$. In this work, we extend beyond this identification and explore the amplitude and the phase landscape as a function of the source parameter space. As an independent and robust observable, the phase exhibits a different dependence on the source properties than the amplitude. In particular, it can acquire a nonzero value even when the source has no intrinsic phase, sourced from the complex nature of the QNM frequencies. Consequently, the phase is strongly dependent on the imaginary part of the QNM frequency, $\omega_n^{\rm Im}$. To explore these aspects we choose the ID to be oscillatory Gaussian pulses, as in Paper~I,
\begin{equation}
\Psi(r_\star)\Big|_{t=0} = A\,\exp\!\Bigg[-\frac{\big(r_\star - r_0\big)^2}{2\sigma^2}\Bigg]\cos\!\Big[\nu\,\big(r_\star - r_0\big)\Big]\,,
\label{eq:id}
\end{equation}
where $A$ (the amplitude) is set to unity. For further discussion on $A$, see Paper~I. The pulse is centered at $r_\star=r_0$, $\sigma$ is the width of the Gaussian envelope and $\nu$ is the carrier frequency. Its Fourier transform is $\Psi(\omega) = \sqrt{2\pi}\sigma\, \tilde{\Psi}(\omega) $ with
\begin{equation}
\tilde\Psi(\omega) =e^{i\omega r_0}\,e^{-\frac{\sigma^2}{2}\big(\omega^2+\nu^2\big)}\,\cosh\big(\sigma^2\omega\nu\big).
\label{eq:fourier_id}
\end{equation}
%
% ----------------------------------------------------------------
% ----------------------------------------------------------------
\begin{figure*}[!ht]
\centering
\includegraphics[width=\linewidth]{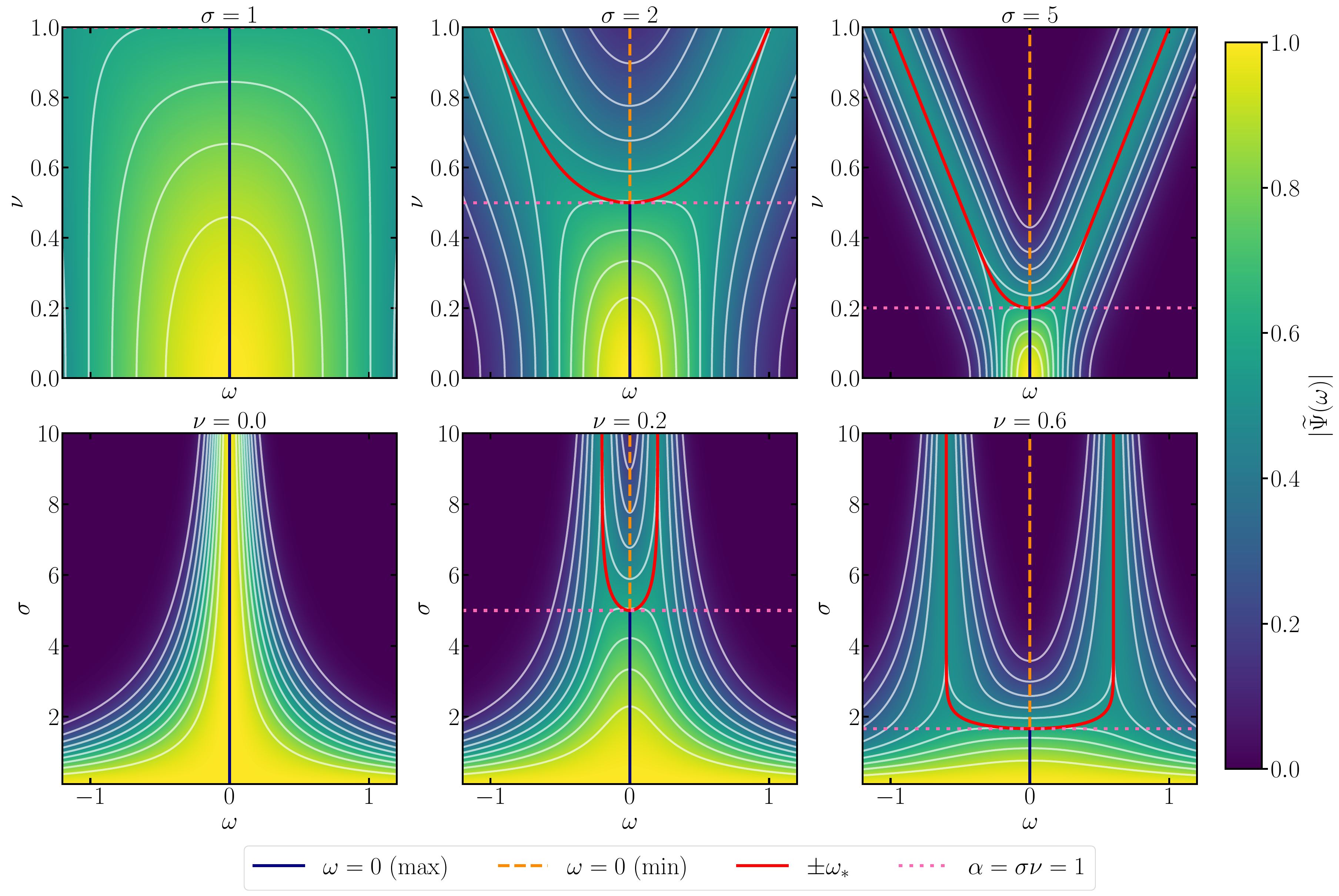}
\caption{\justifying Spectral structure of the initial perturbation in the $(\omega,\nu)$ and $(\omega,\sigma)$ spaces. The top row shows $|\widetilde{\Psi}(\omega)|$ for fixed $\sigma={1,2,5}$, while the bottom row corresponds to fixed $\nu={0,0.2,0.6}$. The color scale represents the spectral amplitude $|\widetilde{\Psi}(\omega)|$, with white contours indicating constant-amplitude levels. The dark blue solid line denotes the central maximum at $\omega=0$ for $\alpha=\sigma\nu<1$, while the orange dashed line marks the corresponding central minimum for $\alpha>1$. The red solid curves indicate the two maxima at $\omega=\pm\omega_\ast$, obtained from $x_\ast=\tanh(\alpha^2x_\ast)$ with $\omega_\ast=\nu x_\ast$. The pink dotted line marks the critical condition $\alpha=\sigma\nu=1$, at which the bifurcation occurs. For $\nu=0$, $\alpha=0$ for all $\sigma$, and no bifurcation is present.}
\label{fig:xstar_approx}
\end{figure*}
% ----------------------------------------------------------------
% ----------------------------------------------------------------
Notice that $|\widetilde{\Psi}(\omega)|$ develops two maxima near $\omega=\pm\nu$. However, for sufficiently small values of $\sigma$, these peaks overlap significantly, eventually merging into a single maximum at $\omega=0$. The transition occurs at the critical value ${\alpha\equiv \sigma\nu=1}$, which marks the bifurcation between the two spectral regimes. 

In Fig.~\ref{fig:xstar_approx} we show $|\widetilde\Psi(\omega)|$ in the $(\omega,\nu)$ plane at fixed $\sigma$ (top row) and at fixed $\nu$ (bottom row).
The source spectrum exhibits a bifurcation at $\sigma\nu\equiv\alpha=1$ (horizontal dotted pink lines). For $\alpha<1$, the spectrum is single-peaked, with a maximum at $\omega=0$. For $\alpha>1$, the overlap between the two branches decreases and the spectrum becomes double peaked and the peak asymptotes to $\omega=\pm\nu$.
For $\sigma=1$ (top left) the pink line showing $\alpha=1$ coincides with the upper edge of the panel. The figure  over the whole demonstrated frequency range the spectrum  shows a single maximum at $\omega=0$ with decreasing amplitude  as $\nu$ (hence $\alpha$) grows. Near $\omega=0$, $|\widetilde\Psi(\omega)|\propto e^{-\alpha^2/2} e^{-\sigma^2(1-\alpha^2)\,\omega^2/2}$, so the peak has width $1/(\sigma\sqrt{1-\alpha^2})$. It flattens as $\alpha\to1$ and splits into two maxima at $\alpha=1$.
At $\alpha=1$, the bifurcation lines $\omega=\pm\nu$ shift to lower carrier frequencies as $\sigma$ increases, reaching $\nu=0.5$ for $\sigma=2$ and $\nu=0.2$ for $\sigma=5$.
The bottom row shows the same transition $|\widetilde\Psi(\omega)|$ as a function of $\sigma$. For $\nu=0$ (bottom-left) the spectrum has a single peak ($\alpha = 0$) at $\omega=0$ that narrows as $1/\sigma$. For $\nu=0.2$ and $\nu=0.6$ the peak splits at $\sigma=5$ and $\sigma=1.67$, respectively, and the two branches converge onto  $\omega=\pm\nu$ as $\sigma$ grows.

These panels define the two source regimes. For $\alpha<1$ the spectral power is concentrated around $\omega=0$: the initial pulse excites dominantly the low frequencies showing a single peak centered at $\omega = 0$, which width scales as $1/\sigma$. On this regime, the ID is spectrally broadbanded and the BH response is mainly driven by its low-frequency content, which also sources the late-time tail (see Paper~I). For $\alpha>1$ the spectral power bifurcates into two branches around $\pm\nu$, and $\nu$ and $\sigma$ become genuine tuning parameters that control the central frequencies and spectral bandwidth, respectively. In this regime, the source effectively acts as a spectral filter of the BH response, allowing specific frequency components to be selectively enhanced or suppressed. As discussed in Sec.~\ref{sec:Tn_structure}, this underlying mechanism governs the selective excitation of individual QNMs.

The spectral bandwidth of each branch also depends on both $\sigma$ and $\nu$. As shown in the top panel of Fig.~\ref{fig:xstar_approx}, above the $\alpha=1$ line, the contours surrounding the red curves become progressively narrower with increasing $\sigma$ at fixed $\nu$. The same trend is observed in the bottom panels. This follows from Eq.~\eqref{eq:fourier_id} where each branch is a Gaussian of width $1/\sigma$ centred at $\pm\nu$, and close to $\alpha=1$ the two overlap. The width of the peak around the maximum, $\Delta\omega=\big[-\partial_{\omega\omega}\ln|\tilde\Psi|\big]^{-1/2}$ is,
\begin{equation}
    \Delta \omega = \frac{1}{\sigma\sqrt{1-\alpha^2\big( 1-x_\ast^2\big)}}\,,
\end{equation}
where $x_\ast$ denotes the location of the maxima in terms of $x\equiv\omega/\nu$, determined by the stationarity condition $x =\tanh(\alpha^2x)$. For $\alpha<1$, the only solution is $x_\ast=0$, corresponding to a single maximum. For $\alpha>1$, three solutions exist: $x=0$, corresponding to a local minimum, and $x=\pm x_\ast$, corresponding to two symmetric maxima, with $x_\ast>0$. At the critical point $\alpha=1$, the denominator vanishes, leading to a formal divergence, signaling the breakdown of the quadratic approximation used to estimate $\Delta\omega$. The existence of this transition at $\alpha=1$ is not specific to the Gaussian envelope: any real modulated pulse $f(r_\star)\cos(\nu r_\star)$ with envelope width $\sigma$ has spectrum $\sim F(\omega-\nu)+F(\omega+\nu)$, where $F$ represents spatial Fourier transform of $f(r_\star)$., which develops two resolved peaks once the branch separation $2\nu$ exceeds the branch width $\sim1/\sigma$. The Gaussian choice only fixes the precise threshold $\alpha=1$ and the pitchfork character of the transition.
%%%%%%%%%%%%%%%%%%%%%%%%%%%%%%%%%%%%%%%%%%%%%%%%%%%%%%%%%%
\subsection{Two-branch interference in the QNM overlap}
%%%%%%%%%%%%%%%%%%%%%%%%%%%%%%%%%%%%%%%%%%%%%%%%%%%%%%%%%%
For the oscillatory Gaussian ID in Eq.~\eqref{eq:id}, the overlap integral Eq.~\eqref{eq:tn_asymptotic} can be evaluated analytically. For purely ingoing ID, using the result of Paper~I\footnote{For purely ingoing ID, $\partial_t\Psi|_{t=0} =\partial_{r_\star}\Psi|_{t=0}$. Integration by parts of the $\partial_{r_\star}\Psi_0$ contribution in the overlap integral, using $\Psi_0 \to 0$ as $r_\star \to \pm\infty$, gives $T_n^{(\partial\Psi)} = T_n^{(\Psi)}$, so that $T_n = 2T_n^{(\Psi)}$. Purely outgoing ID gives $T_n = 0$. See Appendix~A of Paper~I for the full derivation.}, the overlap takes the form of Eq.~\eqref{eq:fourier_id} evaluated at $\omega=\omega_n$, with the two branches written separately and the contribution of the mirror modes $\omega_{n-}=-\omega_{n+}^*$ included as in Eq.~(23) of Paper~I\footnote{The term $e^{i \omega_n r_0}$ has been factored out as in Paper~I.}
\begin{equation}
T_n = -i\sqrt{8\pi}\,\sigma\,\omega_n\Bigg[e^{-\frac{\sigma^2}{2}\big(\omega_n + \nu\big)^2}+ e^{-\frac{\sigma^2}{2}\big(\omega_n - \nu\big)^2}\Bigg]\,.
\label{eq:tn_analytic}
\end{equation}
Equation~\eqref{eq:tn_analytic} makes explicit that the excitation coefficient is the coherent sum of the positive and negative frequency components $\pm \nu$ of the oscillatory source\footnote{Notice that the negative sign of Eq.~\eqref{eq:tn_analytic}, although properly included in our computations, is missing in Eq. (23) of Paper~I. In this work we correct this typo for mathematical consistency.}. The  Gaussian envelope of each branch controls the amplitude of the excitation of the QNM spectrum $|T_n|$, i.e., it sources the resonances and determines the overall resonance strength. The coherent superposition of the two branches, on the other hand, gives rise to interference effects that modify both the QNEC amplitude and phase. 
For $\nu=0$, both of the ID branches merge into one single branch at $\omega=0$, recovering the usual non-oscillatory Gaussian pulse response. In such a limit, the two terms of Eq.~\eqref{eq:tn_analytic} are equal and add in phase.

It is noteworthy that although the closed-form expressions above are specific to the oscillatory Gaussian ID of Eq.~\eqref{eq:id}, the interference mechanism itself is not tied to the Gaussian shape. For a general real radial profile of the ID,
\begin{equation}
\Psi(r_\star)|_{t=0} = f\big(r_\star - r_0\big)\cos\!\Big[\nu\,\big(r_\star - r_0\big)\Big]\,,
\end{equation}
the overlap with a QNM (see Eq.~\eqref{eq:tn_asymptotic}) takes the form,
\begin{equation}
T_n \;\propto\; F\big(\omega_n+\nu\big) + F\big(\omega_n-\nu\big)\,,
\label{eq:generic-overlap}
\end{equation}
where,
\begin{equation}
F(\omega) \equiv \int f\big(r_\star\big)\, e^{i\omega r_\star}\, dr_\star.
\label{eq:F_definition}
\end{equation}
Interference between these two terms, in general, modulates both the amplitude and the phase. The detailed impact of this mechanism for the Gaussian sources is explored in the next sections.

% ----------------------------------------------------------------
\begin{figure}[!ht]
\centering
\includegraphics[width=\linewidth]{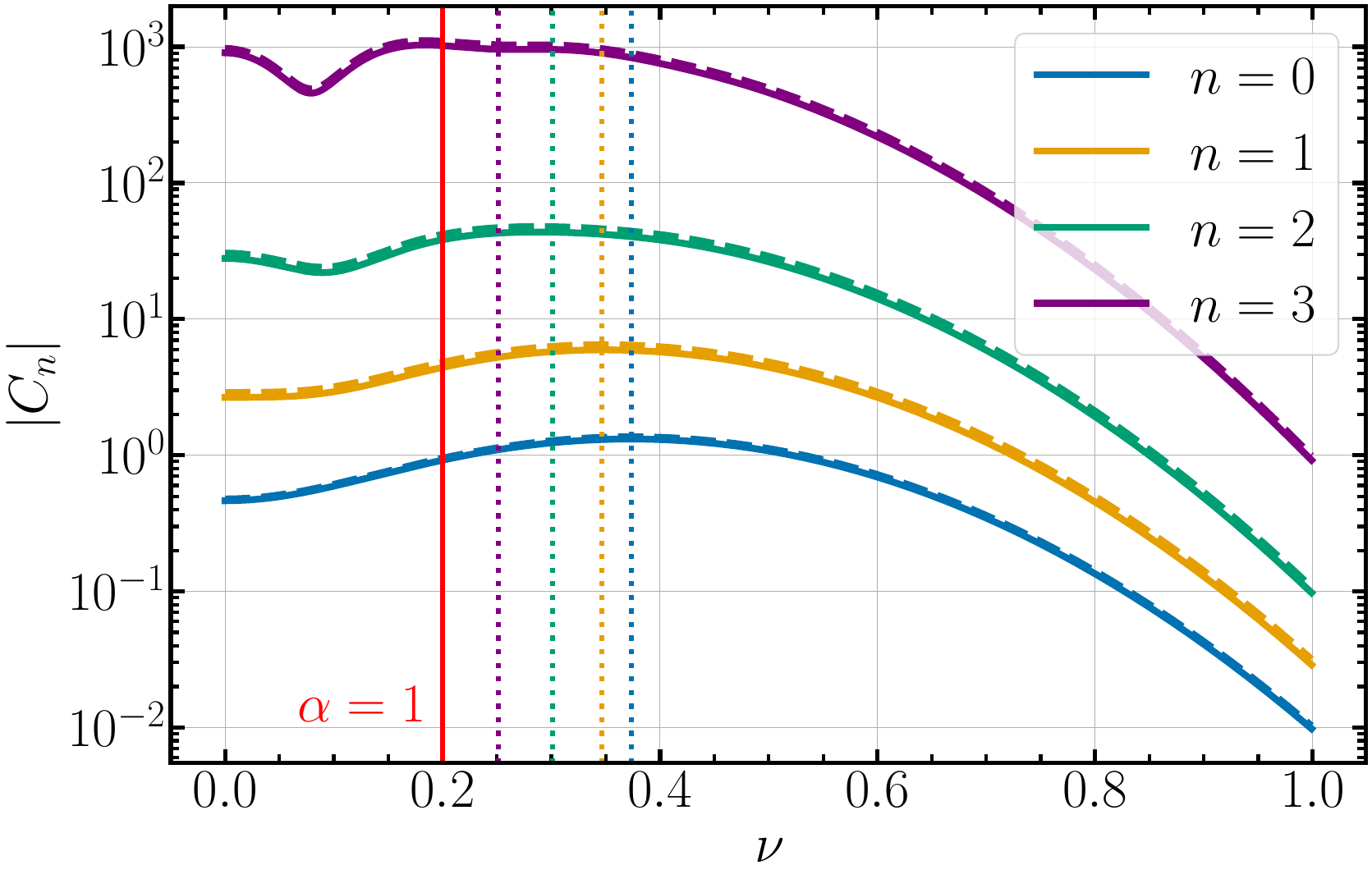}
\caption{\justifying QNECs $|C_n|$ vs carrier frequency $\nu$ at $\sigma = 5$ for the fundamental mode and the first three overtones ($n = 0, 1, 2, 3$, $\ell = 2$). Solid lines: full QNM wavefunctions (Leaver). Dashed lines: asymptotic approximation. Vertical dotted lines mark $\omega_n^{\rm Re}$. For $n=3$ the interference term moves the maxima away from the QNM position. The red vertical line represents $\alpha=1$. The oscillatory modulations at low $\nu$ for $n = 2, 3$ arise from the interference term $\cos\big(2\sigma^2\nu\omega_n^{\rm Im}\big)$. For large $\nu$, exponential suppression dominates. }
\label{fig:qnecs_full}
\end{figure}
%
%%%%%%%%%%%%%%%%%%%%%%%%%%%%%%%%%%%%%%%%%%%%%%%%%%%%%%%%%%
%%%%%%%%%%%%%%%%%%%%%%%%%%%%%%%%%%%%%%%%%%%%%%%%%%%%%%%%%%
\section{Amplitude landscape}
\label{sec:Tn_structure}
%%%%%%%%%%%%%%%%%%%%%%%%%%%%%%%%%%%%%%%%%%%%%%%%%%%%%%%%%%

%%%%%%%%%%%%%%%%%%%%%%%%%%%%%%%%%%%%%%%%%%%%%%%%%%%%%%%%%%
\subsection{Structure of the excitation amplitude}
As QNMs are complex numbers, the overlap in Eq.~\eqref{eq:tn_analytic} contains both an amplitude and a phase. In the current section we focus only on the amplitude.
The squared norm of Eq.~\eqref{eq:tn_analytic} for the ID in Eq.~\eqref{eq:id} is,
\begin{equation}
\big|T_n\big|^2 =16\pi\sigma^2\big|\omega_n\big|^2e^{-\sigma^2\big(\mathcal{P}_n+\nu^2\big)}\mathcal{D}_n \,.
\label{eq:Tn_cosh}
\end{equation}
We have factored out a common Gaussian envelope, $e^{-\sigma^2(\mathcal{P}_n + \nu^2)}$ while leaving a modulation factor $\mathcal{D}_n$ with,
\begin{align}
\mathcal{D}_n \equiv&\, \cosh\Big(2\sigma^2\nu\omega_n^{\rm Re}\Big) + \cos\Big(2\sigma^2\nu\omega_n^{\rm Im}\Big)\,,\label{eq:Dn expression}\\[0.5em]
\mathcal{P}_n \equiv&\, \Big(\omega_n^{\rm Re}\Big)^2 - \Big(\omega_n^{\rm Im}\Big)^2\,. \label{eq:Pn expression}
\end{align}
The exponential factor originates from the spectral envelope of the source. $\mathcal{D}_n$ contains the coherent interaction between the two branches and introduces modulation in the overlap $T_n$ and hence in QNECs. For non-oscillatory Gaussian, $\nu=0$, $\mathcal{D}_n$ is completely independent of the source parameters and is equal to $2$. In that limit, the response is solely determined by the Gaussian envelope. The ID model considered in the current work has reflection symmetry in the carrier frequency across $\nu=0$. Thus the perturbation is invariant under $\nu\rightarrow-\nu$. As a consequence, $T_n(\sigma,-\nu)=T_n(\sigma,\nu)$, and the excitation amplitude is mirror symmetric around $\nu=0$.

Figure~\ref{fig:qnecs_full} shows the QNECs $|C_n|=|B_n \, T_n|$ on a log scale with respect to $\nu\in[0,1]$ for $n=0,1,2,3$ ($\sigma=5$) in different colours. Higher tones have larger amplitude mostly because of the term $e^{+\frac{\sigma^2}{2}(\omega_n^{\rm Im})^2}$ in $|T_n|$, since  $(\omega_n^{\rm Im})$ grows with increasing overtone number $n$. Leaver\footnote{Full QNM wavefunctions are computed via the Leaver series~\cite{leaver:1985ax,Leaver:1986gd}; see Appendix~B of Paper~I for details. The series is truncated at $N_{\rm max} = 300$, sufficient for convergence across all modes and radial ranges considered here.} (solid) and the asymptotic approximation (dashed) overlap throughout.
The $n=0,\,1$ curves rise from $\nu=0$ to a single peak and then falls off. For $n=2,3$ the rise is not monotonic. There is a shallow dip at low $\nu$ before the curve climbs to its peak. This modulation is due to the two-branch interference derived above.
$n=0,\,1,\,2$ peak sits close to $\omega_n^{\rm Re}$ (vertical dotted lines) differing by $-0.25\%$, $+0.21\%$ and $-4.6\%$, respectively. For $n=3$ the dominant maximum is at $\nu \simeq 0.184$, about $27\%$ below $\omega_3^{\rm Re}$. The modulation causes one and two dips respectively for $n=2$ and $n=3$ modes before the exponential term in Eq.~\eqref{eq:Tn_cosh} takes over. The red line marks $\alpha=1$ ($\sigma=5$, $\nu=0.2$), where the source spectrum splits into two branches (see Fig.~\ref{fig:xstar_approx}). 
The modulation factor $\mathcal{D}_n$ is the analytical signature of the coherent interference between the positive- and negative-frequency branches of the source spectrum.
The two terms in $\mathcal{D}_n$ in Eq.~\eqref{eq:Dn expression} exhibit qualitatively distinct behaviors. Since the terms can not simultaneously reach their minima, their sum is strictly positive. 
The behavior of $\mathcal{D}_n$ is controlled by the ratio,
\begin{equation}
\mathfrak{W}_n \equiv \frac{\omega_n^{\rm Re}}{\omega_n^{\rm Im}}\,,
\end{equation}
which distinguishes weakly and strongly damped modes. $\mathcal{P}_n>0$ corresponds to $\mathfrak{W}_n>1$, whereas $\mathcal{P}_n<0$ corresponds to $\mathfrak{W}_n<1$.
In Fig.~\ref{fig:qnecs_full}, for weakly damped modes ($\mathcal{P}_n > 0$), $|C_n|$  has a minimum at $\nu=0$, and it monotonically increases before exponential damping due to the Gaussian envelope in Eq.~\eqref{eq:Tn_cosh} takes over. On the other hand, for strongly damped modes ($\mathcal{P}_n<0$), $\nu=0$ is instead a local maximum. The oscillatory contribution can then produce additional minima at finite $\nu$ before the hyperbolic term dominates at larger frequency. These modulations are controlled by the ratio $\mathfrak{W}_n$ and leads to a local suppression of the excitation amplitude for overdamped modes at finite $\nu$, visible as the dips in the $C_2$ and $C_3$ curves of Fig.~\ref{fig:qnecs_full}. These modulations can further be understood from extrema analysis of the overlap and is provided in the next sections.

%%%%%%%%%%%%%%%%%%%%%%%%%%%%%%%%%%%%%%%%%%%%%%%%%%%%%%%%%%
\subsection{Stationary conditions and spectroscopic structures}
\label{sec:Stationary conditions}
%%%%%%%%%%%%%%%%%%%%%%%%%%%%%%%%%%%%%%%%%%%%%%%%%%%%%%%%%%
In this section, we identify the ID configurations that maximize the excitation of a given QNM. We do so by analysing the extrema curves of the overlap in the $(\sigma,\nu)$ parameter space.
Since the logarithm is monotonic, the stationary points of the excitation amplitude can be obtained from $\ln|T_n|^2$.
The stationarity conditions $(\partial_{\nu,\sigma} \ln{|T_n|^2} = 0)$ are found to be respectively,
\begin{align}
\mathcal{S}_n &= \nu\,\mathcal{D}_n\,,
\label{eq:extremum_nu}\\[0.5em]
\sigma^2\Bigg[\mathcal{P}_n+\nu^2-2\nu\,\frac{\mathcal{S}_n}{\mathcal{D}_n}\Bigg]&=1\,,
\label{eq:extremum_sigma}
\end{align}
where,
\begin{equation}
\mathcal{S}_n \equiv \omega_n^{\rm Re}\sinh\Big(2\sigma^2\omega_n^{\rm Re}\nu\Big)
- \omega_n^{\rm Im}\sin\Big(2\sigma^2\omega_n^{\rm Im}\nu\Big).
\label{eq:S_definition}
\end{equation}
Equation~\eqref{eq:extremum_nu} determines the  stationary points in $\nu$ at fixed $\sigma$, whereas Eq.~\eqref{eq:extremum_sigma} determines the stationary points in $\sigma$ at fixed $\nu$. A genuine stationary point in the full $(\sigma,\nu)$ plane must satisfy both conditions simultaneously.

The $\nu=0$ satisfies Eq.~\eqref{eq:extremum_nu} for any values of $n$. The $\nu\nu$ component of the Hessian ($H_{\nu\nu}$) at $\nu=0$ is,
\begin{equation}
H_{\nu\nu}=\partial_{\nu\nu}\ln\big|T_n\big|^2\bigg|_{\nu=0} = 2\sigma^2\Big(\sigma^2\mathcal{P}_n - 1\Big),
\label{eq:hessian_nu}
\end{equation}
whose sign identifies a critical width $\sigma^*_n = 1/\sqrt{\mathcal{P}_n}$ for $\mathcal{P}_n > 0$, coinciding with the optimal pure-Gaussian width $\sigma^*_n$ identified in Paper~I. At this point $H_{\nu\nu}$ vanishes.
For $\sigma<\sigma_n^*$, $H_{\nu\nu}<0$ and $\nu=0$ is a local maximum in the $\nu$ direction. For $\sigma>\sigma_n^*$, $H_{\nu\nu}>0$ and it becomes a local minimum. Fig.~\ref{fig:qnecs_full} shows this feature for specific choice of $\sigma=5$. Here and throughout, maxima and minima refer to local extrema with respect to one parameter ($\sigma$ or $\nu$) while holding the other fixed. Stationary points in the full $(\sigma,\nu)$ plane are discussed in Sec.~\ref{sec:lorentzian}. 

Evaluating the second derivative along the $\sigma$ direction at the critical point $(\nu,\sigma) = (0,\sigma^*_n)$  provides
\begin{equation}
\partial_{\sigma\sigma}\ln\big|T_n\big|^2\bigg|_{(0,\sigma^*_n)}
= -4\mathcal{P}_n < 0 \quad (\mathcal{P}_n > 0),
\end{equation}
Thus, for $\mathcal{P}_n>0$, the point $(0,\sigma_n^*)$ is a maximum along the $\sigma$ direction while its second derivative along $\nu$ vanishes.
For $\mathcal{P}_n < 0$ ($n \geq 2$ for $\ell = 2$), Eq.~\eqref{eq:hessian_nu} gives $\partial_{\nu\nu}\ln|T_n|^2 < 0$ for all $\sigma$, so $\nu = 0$ remains a local maximum in the $\nu$ direction. 
%
%%%%%%%%%%%%%%%%%%%%%%%%%%%%%%%%%%%%%%%%%%%%%%%%%%%%%%%%%%
\subsection{Carrier frequency extrema }
\label{subsec:nu_maxima}
%%%%%%%%%%%%%%%%%%%%%%%%%%%%%%%%%%%%%%%%%%%%%%%%%%%%%%%%%%
%
\begin{figure}[!ht]
    \centering
    \includegraphics[width=1\linewidth]{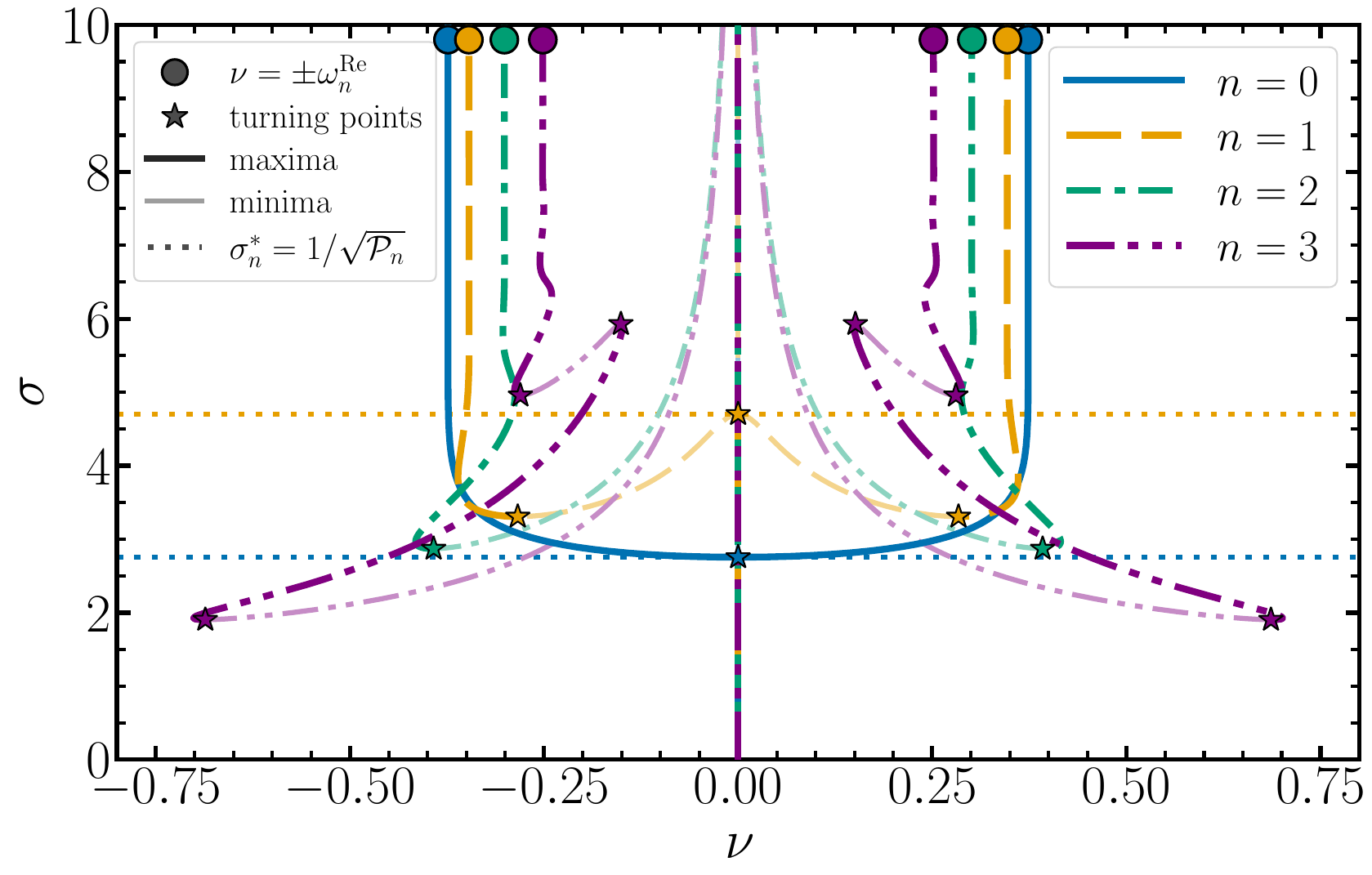}
       \caption{\justifying Stationary points of the excitation amplitude $|C_n|$ along $\nu$, Eq.~\eqref{eq:extremum_nu}, in the $(\nu,\sigma)$ plane for the first four $\ell=2$ QNMs. Dark curves are maxima and light curves are minima. The line $\nu=0$ is stationary for every mode: for $n=0,1$ it is a maximum below the critical width $\sigma_n^*$ (horizontal dotted lines) and a minimum above it, while for $n=2,3$ it is a maximum at all widths. The circles on the upper axis mark $\nu=\pm\omega_n^{\rm Re}$. The star markers show each curve's turning points.}
    \label{fig:numax_vs_sigma}
\end{figure}
In Fig.~\ref{fig:numax_vs_sigma} we show all extremas, computed analytically,  of the excitation amplitude $|C_n|$  along the $\nu$ direction as a function of the width for the first four $\ell=2$ QNMs using Eq.~\eqref{eq:extremum_nu}. 
Maxima are shown in dark colours and minima in light ones. The circles indicate the points at which  $\nu=\omega_n^{\rm Re}$. For $n=0,1$ the critical widths $\sigma_n^*$ are indicated by the horizontal dotted lines. As discussed before, the vertical line $\nu=0$ is an extrema for all the modes. 
For the fundamental mode (blue) it is a maximum below $\sigma_0^*\simeq2.75$ and a minimum above it, and at $\sigma_0^*$ two maxima appear from $\nu=0$ and move outwards with increasing $|\nu|$, before asymptoting towards the circles at $\nu=\pm\omega_0^{\rm Re}$.
For the first overtone (yellow), $\nu=0$ changes from a maximum to a minimum at $\sigma_1^*\simeq4.7$. Unlike the fundamental mode, the $\nu\neq0$ branch of the extrema curve transitions into a minima and moves towards smaller widths until it reaches a turning point (star marked) at $\sigma\simeq3.3$ and $\nu=0.28$ where it becomes a maximum that progresses towards $\nu=\pm\omega_1^{\rm Re}$. This produces an interval of widths $3.3<\sigma<4.7$ in which the central $(\nu=0)$ and the off-centered $(\nu\neq0)$ maxima coexist.

For $n=2$ and $n=3$ (green and purple, respectively), no critical width exists and $\nu=0$ remains a local maximum for all $\sigma$. As in the $n=1$ case, the extrema near $\nu=0$ are minima for $\nu\neq0$. However, unlike $n=1$ case, they approach $\nu=0$ in the limit $\sigma\to\infty$. As $\nu$ increases, the curves move towards smaller $\sigma$, reaching turning points (star marked) at $(\nu,\sigma)\simeq(0.39,2.87)$ and $(0.69,1.90)$ for $n=2$ and $n=3$, respectively. Beyond these points, both curves turn towards increasing $\sigma$ and decreasing $\nu$. For $n=2$, the curve remains a maximum and approaches $\nu=\omega_2^{\rm Re}$, passing through a bend at $(\nu,\sigma)\simeq(0.29,4.95)$. In contrast, the $n=3$ curve undergoes two additional turning points at $(\nu,\sigma)\simeq(0.15,5.93)$ and $(0.28,4.96)$, where the extrema flips to minima and maxima respectively, before approaching $\nu=\omega_3^{\rm Re}$ at large $\sigma$. The structure for negative $\nu$ is identical under the reflection $\nu\to-\nu$.

The transition of $\nu=0$ from a local maximum to a local minimum can be  understood by expanding Eq.~\eqref{eq:Tn_cosh} around $\nu=0$,
\begin{align}
\ln|T_n|^2&\simeq\ln|T_n|^2\Big|_{\nu=0}-\sigma^2\nu^2+\sigma^4\mathcal{P}_n\nu^2\notag\\[0.5em].
\label{eq:Tn_small_nu}
\end{align}
The first term, $-\sigma^2\nu^2$, comes from the Gaussian factor $e^{-\sigma^2\nu^2}$ and lowers the amplitude as $|\nu|$ increases. The second term, $\sigma^4\mathcal{P}_n\nu^2$, arises from $\mathcal{D}_n$, which contains competing hyperbolic and trigonometric contributions associated with $\omega_n^{\rm Re}$ and $\omega_n^{\rm Im}$, respectively. For weakly damped modes ($\mathcal{P}_n>0$), the hyperbolic contribution dominates, making the second term positive which competes with the Gaussian suppression, determining whether $\nu=0$ is a local maximum or minimum. For narrow pulses, i.e. $\sigma<\sigma_n^*=1/\sqrt{\mathcal{P}_n}$, $\nu=0$ becomes a maximum, while for $\sigma>\sigma_n^*$ the term from $\mathcal{D}_n$ is larger. Thus $\nu=0$ becomes a minimum and the maxima shifts continuosly from $\nu=0$ to $\nu=\pm\omega_n^{\rm Re}$ for $n=0$ and goes through two turning points for $n=1$ (see Fig.~\ref{fig:numax_vs_sigma}). For strongly damped modes ($\mathcal{P}_n<0$) both terms lower the amplitude, and $\nu=0$ remains a maximum for all the $\sigma$ range considered.

Both $\mathcal{S}_n$ and $\mathcal{D}_n$ in Eq.~\eqref{eq:extremum_nu} contain hyperbolic and trigonometric functions. For $\nu>0$, the hyperbolic terms grow monotonically, while the trigonometric terms oscillate between positive and negative values. The resulting extrema therefore depend on the phase $2\sigma^2\omega_n^{\rm Im}\nu$, with $\omega_n^{\rm Im}$ setting the oscillation scale. These oscillations reflect the interference phenomena discussed above.
For example, the maximum value of $\nu$ at which an extremum changes from a maximum to a minimum, or vice versa, depends on the mode, as shown in the figure. The number of turning points in the outer branch within a given $\sigma$ range also increases with the damping rate. This trend can be partly understood from the factors multiplying the hyperbolic and trigonometric terms in $\mathcal{S}_n$, namely $\omega_n^{\rm Re}$ and $\omega_n^{\rm Im}$, respectively. As the damping rate increases, the relative contribution of the oscillatory trigonometric term is enhanced, limiting the dominance of the monotonically growing hyperbolic term.
This also explains the different behavior of the $n=0$ and $n=1$ modes near $\sigma_n^*$. For $n=0$, the hyperbolic term dominates, and the off-centre maxima emerges continuously from $\nu=0$ and transitions to $\nu=\pm \omega_{n}^{\rm Re}$. For $n=1$, the stronger damping enhances the strength of the interference term, leading to the appearance of an additional branch of minima and a turning point.

The stationary structure also shows that efficient QNM excitation does not generally require direct resonance matching between the source carrier frequency and the QNM oscillation frequency. At finite and sufficiently small $\sigma$, the $\nu$-maximum of the excitation amplitude can lie away from  $\nu=\omega_n^{\rm Re}$. For $\sigma<\sigma_n^*$ when $\mathcal{P}_n>0$, $\nu=0$ becomes a maximum. In this regime, the source is sufficiently broadband that its spectral content overlaps with the QNM even without a carrier frequency tuned to the mode. Conversely, as $\sigma$ increases and the source becomes narrower in frequency, the off-centre maxima move towards $\nu=\pm\omega_n^{\rm Re}$. The familiar resonance-matching picture is therefore recovered only in the narrowband limit. This distinction is relevant when interpreting the excitation of individual modes by perturbations whose spectral content is broad compared to the QNM linewidth.

For higher overtones the damping induced modulation introduces further features. At $\sigma\simeq3.05$ for $n=2$ and at $\sigma\simeq2.87$ and $5.06$ for $n=3$ the overlap contains minima at $\nu=\omega_n^{\rm Re}$. These are the points where the light curves of $n=2$ and $n=3$ in Fig.~\ref{fig:numax_vs_sigma} cross $\nu=\omega_n^{\rm Re}$, marked by the circles on the upper axis. Thus, depending on the spectral width of the sources, QNMs with large damping can be minimally excited despite the carrier frequency satisfying the resonance conditions $\nu=\omega_n^{\rm Re}$.

This suggests that QNM excitation by realistic astrophysical sources need not require a direct matching between orbital and QNM frequencies when the source has sufficiently broad spectral support. The implications of this mechanism for astrophysical sources will be explored in future work.

%%%%%%%%%%%%%%%%%%%%%%%%%%%%%%%%%%%%%%%%%%%%%%%%%%%%%%%%%%
\subsection{Lorentzian structure of the width-extremization}
\label{sec:lorentzian}
%%%%%%%%%%%%%%%%%%%%%%%%%%%%%%%%%%%%%%%%%%%%%%%%%%%%%%%%%%

In this section, we investigate the complementary question: for a given carrier frequency $\nu$, which pulse widths maximally and minimally excite a QNM? At fixed $\nu$ and for $n=0,1$, an optimal width exists only if the carrier frequency lies outside the linewidth of the mode, $|\nu-\omega_n^{\rm Re}|>\omega_n^{\rm Im}$. For $n=2,3$, however, $|\nu-\omega_n^{\rm Re}|>\omega_n^{\rm Im}$ is only satisfied at sufficiently large $\sigma$. The power spectrum of a damped mode is a Lorentzian centred at $\omega_n^{\rm Re}$ with half width $\omega_n^{\rm Im}$, so this condition places $\nu$ outside its linewidth. 
These aspects can be understood by first noting that Eq.~\eqref{eq:extremum_sigma} is invariant under $\nu\to-\nu$, so the positive and negative $\nu$ domains exhibit identical behavior. In either domain, the transcendental equation admits two solution branches for $\mathcal{P}_n>0$ and one branch for $\mathcal{P}_n<0$. As discussed below, when they exist, one lies in the range $0\leq|\nu|<\omega_n^{\rm Re}$, while the other lies in $|\nu|>\omega_n^{\rm Re}$, for the tones considered here. We refer to these as the \textit{inner} and \textit{outer} branches, respectively. 
\begin{figure}[!ht]
    \centering
    \includegraphics[width=1\linewidth]{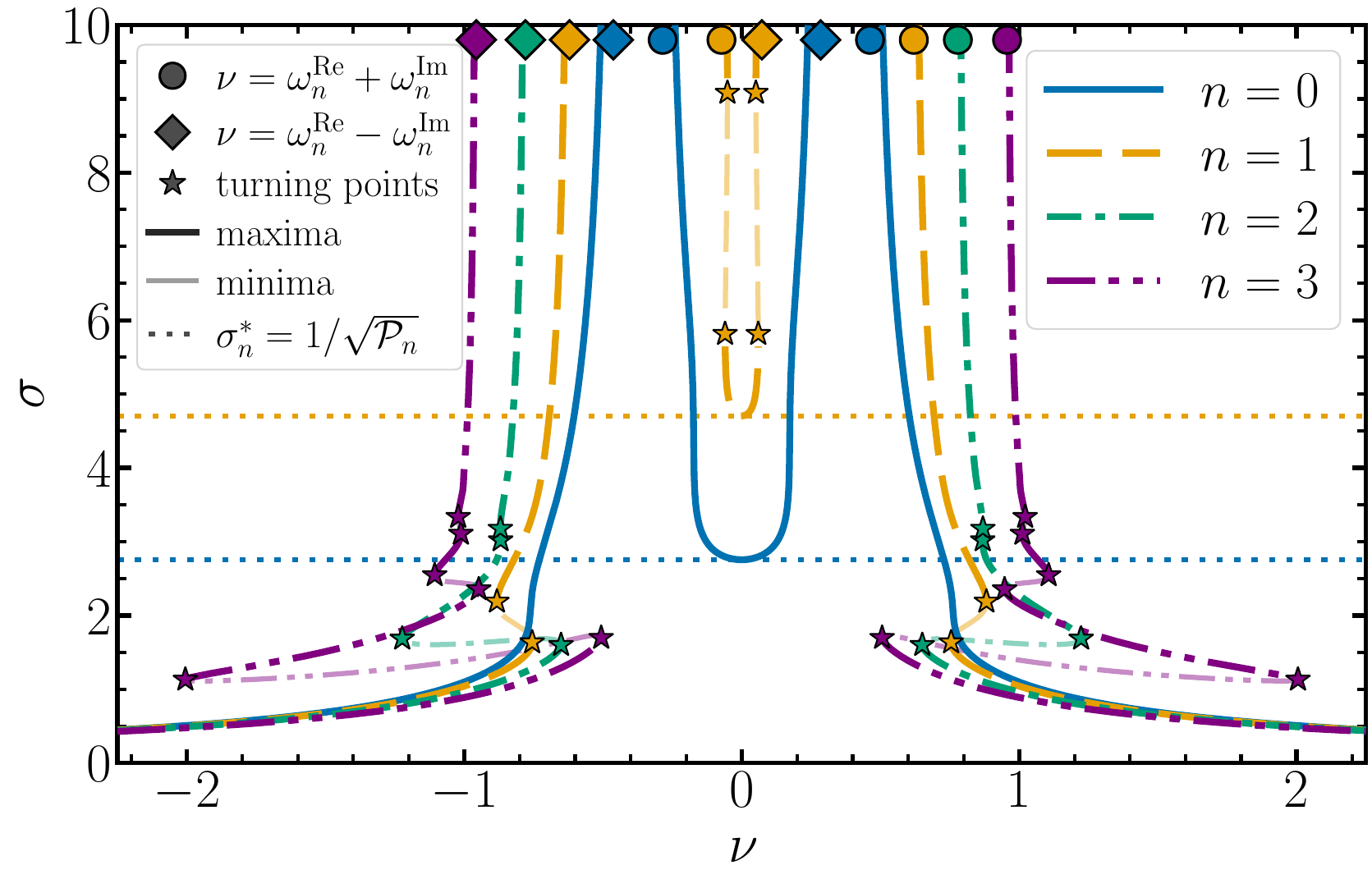}
       \caption{\justifying Maxima (dark) and minima (light) of $|C_n|$ along $\sigma$ for $n=0,1,2,3$ is shown. $\sigma_n^*$ for $n=0,1$ is shown as dotted lines. The markers on the top axis are the large-$\sigma$ limits of $\nu_\pm$, Eq.~\eqref{eq:nu_branches}. For $\nu>0$ they are computed with the mode $\omega_{n+}$, with circles at $\omega_n^{\rm Re}+\omega_n^{\rm Im}$ and diamonds at $\omega_n^{\rm Re}-\omega_n^{\rm Im}$, while for $\nu<0$ they are computed with its mirror mode $\omega_{n-}=-\omega_{n+}^*$. The star markers show each curve's turning points. }
    \label{fig:sigmamax}
\end{figure}
For a fixed $\nu>0$ and sufficiently large $\sigma$ with $2\sigma^2\nu\omega_n^{\rm Re}\gg1$, the hyperbolic contribution to $\mathcal{D}_n$ dominates. In this limit, Eq.~\eqref{eq:extremum_sigma} reduces to,
\begin{equation}
\Big(\nu - \omega_n^{\rm Re}\Big)^2 = \Big(\omega_n^{\rm Im}\Big)^2 + \frac{1}{\sigma^2}.
\label{eq:sigma_branch_large}
\end{equation}
Solving for the carrier frequency gives the asymptotic branches,
\begin{equation}
\nu_{\pm} = \omega_n^{\rm Re} \pm \sqrt{\Big(\omega_n^{\rm Im}\Big)^2  + \frac{1}{\sigma^2}},
\label{eq:nu_branches}
\end{equation}
which approach,
\begin{equation}
\nu_{\pm}\longrightarrow\omega_n^{\rm Re}\pm\omega_n^{\rm Im}\qquad\big(\sigma\rightarrow\infty\big).
\end{equation}
Thus $\nu_+$ and $\nu_-$ are the asymptotic limit of the outer and inner branch respectively.
Equivalently, Eq.~\eqref{eq:sigma_branch_large} gives the asymptotic stationary width at fixed carrier frequency,
\begin{equation}
\sigma_{\rm opt}(\nu) \approx \frac{1}{\sqrt{ 
\Big(\nu-\omega_n^{\rm Re}\Big)^2 - \Big(\omega_n^{\rm Im}\Big)^2}},
\label{eq:sigma_opt_lorentzian}
\end{equation}
real for $|\nu-\omega_n^{\rm Re}|>\omega_n^{\rm Im}$, and diverging as $\nu\to\omega_n^{\rm Re}\pm\omega_n^{\rm Im}$. The asymptotic behavior of $\nu_{\pm}\sim\omega_n^{\rm Re}\pm\omega_n^{\rm Im}$ are the half-power points of a Lorentzian of half width $\omega_n^{\rm Im}$ centered at $\omega_n^{\rm Re}$. For $|\nu-\omega_n^{\rm Re}|>\omega_n^{\rm Im}$, Eq.~\eqref{eq:sigma_opt_lorentzian} gives a finite optimal width, $\sigma_{\rm opt}$, which diverges at $\nu_{\pm}$, whereas for $|\nu-\omega_n^{\rm Re}|<\omega_n^{\rm Im}$ the amplitude grows with $\sigma$ and no finite optimal width exists.
For $|\nu|\ll \omega_n^{\rm Re}-\omega_n^{\rm Im}$  the width $\sigma_{\rm max}$ that maximizes the amplitude follows,
\begin{equation}
\sigma_{\rm max}(\nu) = \sigma_n^* + \frac{1}{2}\big(\sigma_n^*\big)^3\,\nu^2 + \mathcal{O}\big(\nu^4\big)\,.
\label{eq:sigma_max_expansion}
\end{equation}
The expansion above connects $(\nu,\sigma)=(0,\sigma_n^*)$ and $\nu=\omega_n^{\rm Re}-\omega_n^{\rm Im}$ through the inner branch. Another second branch occurs above $\nu=\omega_n^{\rm Re}+\omega_n^{\rm Im}$, as can be seen from Eq.~\eqref{eq:sigma_opt_lorentzian}.

Figure~\ref{fig:sigmamax} shows the inner ${0\leq|\nu|<\omega_n^{\rm Re}}$ and outer ${|\nu|>\omega_n^{\rm Re}}$ extrema branches, where the dark parts of the extrema are maxima and light ones are minima as in Fig.~\ref{fig:numax_vs_sigma}. Here, the outer branch  emerges from large $|\nu|$ and small $\sigma$, bend towards the center as $\sigma$ increases, and  asymptotes towards the circles through the $\nu_+$ branch. These vertical ends are the $\sigma_{\rm opt} \rightarrow \infty$  limit of Eq.\eqref{eq:sigma_opt_lorentzian} at $|\nu|=|\omega_n^{\rm Re}|+\omega_n^{\rm Im}$.
Starting at $\nu=0$ and $\sigma=\sigma_n^*$, the inner branches (when exists) asymptote towards the diamonds (circles) for $\nu>0\, (\nu<0)$ through the $\nu_-$ branch. We find the inner branches to exist only for $n=0,1$. The inner branches of the positive and negative $\nu$ domain form an U shape which meets at $(0,\sigma_n^*)$.  The asymmetry is only in the markers: for $\nu<0$ they represent the linewidth of the mirror mode. Consequently, circles and diamonds swap places between the inner and the outer branches relative to the $\nu>0$ domain. 

For ${n=0}$ (blue) the outer branch is a maximum everywhere. 
For $n=1$ (yellow) the extrema starts as a maximum at large $\nu$ low and $\sigma$, till it reaches a turning point (star markers) at $(\nu,\sigma)\simeq(0.76,1.64)$, where it becomes a minimum and moves outwards until a second turning point at $(0.88,2.19)$. Beyond this point, the stationary point becomes a local maximum once again, with its position approaching $\nu=\omega_1^{\rm Re}+\omega_1^{\rm Im}$. For $n=2$ (green) the first stretch of minima is almost horizontal, between the turning points $(0.65,1.60)$ and $(1.22,1.69)$, and a second, very short one appears near $\nu\simeq0.87$ for $3.02<\sigma<3.18$. For $n=3$ (purple) the first minima portion is  longer, from $(0.51,1.70)$ to $(2.01,1.13)$, and it is followed by two short ones near $\nu\simeq1$, for $2.36<\sigma<2.55$ and $3.11<\sigma<3.34$. In these intervals, at fixed $\nu$, the amplitude has more than one maximum along $\sigma$. As in Fig.~\ref{fig:numax_vs_sigma}, the number of turning points grows with the damping of the mode. A similar structure exists in the $\nu<0$ domain. 

The turning points of these curves come from the interference between the two branches of the source as shown in Sec.~\ref{subsec:nu_maxima}. Both $\mathcal{D}_n$ and $\mathcal{S}_n$ depend on the width and the carrier frequency only through $\sigma^2\nu \,\omega_n^{\rm Re}$ and $\sigma^2\nu \,\omega_n^{\rm Im}$, and $\mathcal{S}_n$ enters Eq.\eqref{eq:extremum_sigma} through the ratio $\mathcal{S}_n/\mathcal{D}_n$. When the hyperbolic parts dominate, $\mathcal{S}_n/\mathcal{D}_n\to\omega_n^{\rm Re}$ and we recover the Lorentzian limit of Eq.\eqref{eq:sigma_branch_large}. At  smaller $\sigma^2\nu\,\omega_n^{\rm Re}$ the trigonometric terms, which carry the interference modulation of $\mathcal{D}_n$, make this ratio oscillate with the phase $\sigma^2\nu\,\omega_n^{\rm Im}$, and these oscillations can turn a maximum into a minimum and back. This is the same mechanism that causes the turning points in Fig.~\ref{fig:numax_vs_sigma}.

A stationary point in the full $(\sigma,\nu)$ plane must satisfy Eqs.~\eqref{eq:extremum_nu} and~\eqref{eq:extremum_sigma} simultaneously. Combining both conditions gives the necessary condition,
\begin{equation}
\nu^2 = \mathcal{P}_n - \frac{1}{\sigma^2}.
\label{eq:joint extrema constraint}
\end{equation}
For real $\nu$, this requires $\mathcal{P}_n>0$ and $\sigma\geq\sigma_n^*=1/\sqrt{\mathcal{P}_n}$ The equality corresponds to the point $(\nu,\sigma)=(0,\sigma_n^*)$, while other solutions require $\sigma>\sigma_n^*$.
Eq.~\eqref{eq:joint extrema constraint} defines a candidate locus for joint stationarity in the $(\sigma,\nu)$ plane. A point on this locus is a joint stationary point only if it also satisfies Eq.~\eqref{eq:extremum_nu} (Eq.~\eqref{eq:extremum_sigma} then holds automatically). Thus, this locus exists only for the weakly damped modes and is absent otherwise.
For $\mathcal{P}_n>0$ we find only $(0,\sigma_n^*)$ satisfies both the extrema conditions. At this point, $\partial_{\nu\nu}\ln|T_n|^2$ and $\partial_{\nu\sigma}\ln|T_n|^2$ vanish, so the determinant of the Hessian is zero and the second-order test does not decide.

Its character follows from the next order. Along the $\sigma$-maximum branch of Eq.~\eqref{eq:sigma_max_expansion},
\begin{equation}
\ln\big|T_n\big|^2\Big|_{\sigma_{\rm max}(\nu)}-\ln\big|T_n\big|^2\Big|_{(0,\sigma_n^*)}\simeq\frac{\big(\omega_n^{\rm Re}\big)^4+\big(\omega_n^{\rm Im}\big)^4}{3\,\mathcal{P}_n^4}\,\nu^4>0 .\end{equation}
Therefore $(0,\sigma_n^*)$ is not a maximum in the full plane: the excitation keeps growing along the $\sigma$-maximum branch that starts there.

\begin{figure*}[!ht]
\centering
\includegraphics[width=\textwidth]{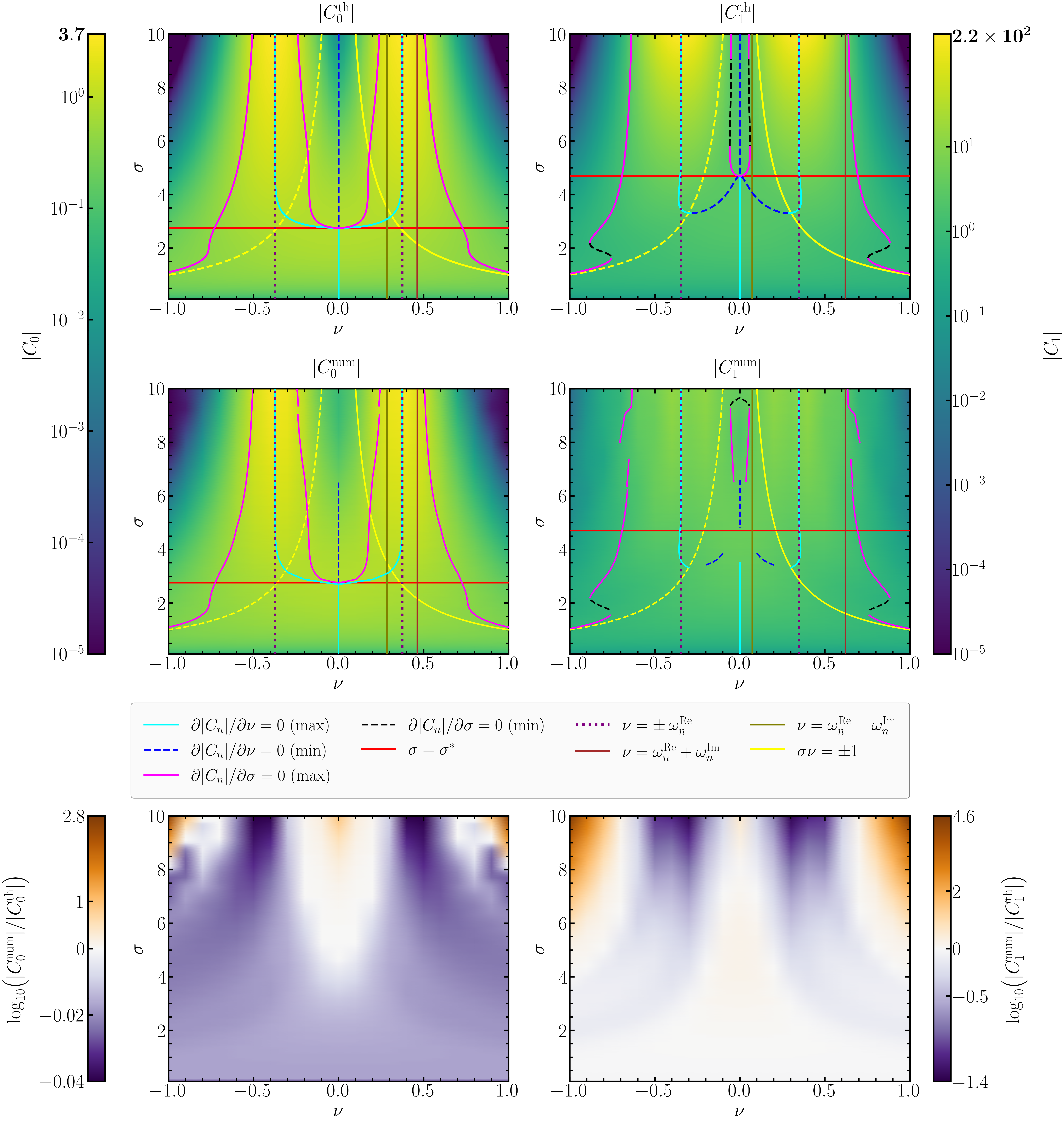}
\caption{\justifying From top to bottom, the rows show the theoretical prediction, the numerical result obtained from fitting the RW numerical data, and the difference $\log_{10}(|C_n^{\rm num}|/|C_n^{\rm th}|)$, with its own colour scale for each mode. In each row, the left panel corresponds to $n=0$ and the right panel to $n=1$. Cyan solid and blue dashed curves are maxima and minima along $\nu$. Magenta solid and black dashed curves are maxima and minima along $\sigma$. The red horizontal line is $\sigma=\sigma_n^{*}$, the purple dotted lines are $\nu=\pm\omega_n^{\rm Re}$, the brown and olive lines are $\nu=\omega_n^{\rm Re}+\omega_n^{\rm Im}$ and $\nu=\omega_n^{\rm Re}-\omega_n^{\rm Im}$, and the yellow curves are $\sigma\nu=\pm1$. }
\label{fig:Cn_heatmaps}
\end{figure*}

%%%%%%%%%%%%%%%%%%%%%%%%%%%%%%%%%%%%%%%%%%%%%%%%%%%%%%%%%%
\subsection{Numerical validation of the amplitude landscape}
%%%%%%%%%%%%%%%%%%%%%%%%%%%%%%%%%%%%%%%%%%%%%%%%%%%%%%%%%%
In Fig.~\ref{fig:Cn_heatmaps} we provide a global view of the QNM excitation landscape $C_n$ in the source-parameter space $(\sigma, \nu)$. For the fundamental mode ($n=0$, left) and first overtone ($n=1$, right), the analytical prediction $|C_n^{\rm th}|$ for the QNECs obtained using Eq.~\eqref{eq:tn_analytic} is shown in the top row and compared directly with the QNECs extracted from time-domain RW evolutions $|C_n^{\rm num}|$ in the middle row.
The numerical evolutions are performed at fixed $r_0=100$ and $r_\star^{\rm obs}=100$ using the RW solver of Paper~I, while $\sigma$ and $\nu$ are varied over the parameter ranges shown in the figure.
The bottom row shows $\log_{10}(|C_n^{\rm num}|/|C_n^{\rm th}|)$ and quantifies their agreement across the full parameter space. Beyond the amplitude itself, the overlaid curves show the stationary structure derived in Eq.~\eqref{eq:extremum_nu} and \eqref{eq:extremum_sigma}. In particular, both the theoretical and numerical heatmaps display the source bifurcation $\alpha=\sigma|\nu|=1$, the mode dependent critical width $\sigma_n^*$, the resonant frequencies $\nu=\pm \omega_n^{\rm Re}$ and the asymptotic Lorentzian boundaries $\nu=\omega_n^{\rm Re}\pm\omega_n^{\rm Im}$.

The cyan solid and blue dashed curves show the solutions of Eq.~\eqref{eq:extremum_nu}, corresponding respectively to maxima and minima of the excitation amplitude along the $\nu$ direction at fixed $\sigma$. For both $n=0$ and $n=1$, the branch $\nu=0$ is a local maximum for $\sigma<\sigma_n^*$, as discussed in Sec.~\ref{sec:Stationary conditions}. At $\sigma_0^*\simeq2.75$ and $\sigma_1^*\simeq4.70$ (red solid horizontal lines) the stationary structure changes such that for $\sigma>\sigma_n^*$, $\nu=0$ becomes a local minimum while two symmetric maxima appear at finite $\nu$. The magenta solid and black dashed curves show the solutions of Eq.~\eqref{eq:extremum_sigma}, corresponding to maxima and minima of the excitation amplitude along the $\sigma$ direction at fixed $\nu$. Their large $\sigma$ behaviour comes from the Lorentzian structure derived in Sec.~\ref{sec:lorentzian} where the stationary branches approach the asymptotic frequencies $\nu=\omega_n^{\rm Re}\pm\omega_n^{\rm Im}$. These limits are shown by the vertical brown and olive curves and in between the interval no finite $\sigma$ satisfies the stationarity condition.
The yellow curve $\alpha=\sigma\nu=\pm1$ marks the spectral bifurcation of the source itself which is independent of the QNMs and therefore identical in both panels, theoretical and numerical.

The middle row provides a direct time-domain validation of this stationary structure described previously. The cyan/blue and magenta/black branches are extracted independently from the fitted amplitudes, by computing and finding numerically the extrema conditions obtained in Secs.~\ref{subsec:nu_maxima} and~\ref{sec:lorentzian}. Their agreement with the theoretical branches in the top row shows that both the location and the change of character of the stationary points are mostly recovered by the numerical evolutions. The QNEF $B_n$ being independent of $(\sigma,\nu)$, does not shift the stationary branches, whose geometry is determined by the source overlap integral $T_n$ and hence by the complex QNM frequency.
The agreement is particularly clear for the fundamental mode, while the first overtone displays larger fluctuations in the extracted branches at large $\sigma$ and low $\nu$. The bottom row quantifies this comparison through $\log_{10}(|C_n^{\rm num}|/|C_n^{\rm th}|)$, which remains close to zero over most of the parameter space.

The exceptions are the upper corners, $\sigma\gtrsim8$ and $|\nu|\gtrsim0.85$, where $|C_n^{\rm th}|$ drops to about $10^{-7}$ or below and the fitted amplitude stays up to about three ($n=0$) and five ($n=1$) orders of magnitude above it. For $n=1$, and in the band $\sigma\gtrsim6$, $0.2\lesssim|\nu|\lesssim0.6$, the extracted amplitude is between about $1.5$ and $6$ times smaller than the theory for $\sigma<8$, and up to about $20$ times for $\sigma\gtrsim9$.
The numerical landscape also recovers the expected $\nu\rightarrow-\nu$ symmetry, providing an additional consistency check of the time-domain extraction. The median of $|\log_{10}(|C_n^{\rm num}|/|C_n^{\rm th}|)|$ is $0.019$ for $n=0$ and $0.134$ for $n=1$ over the whole plane, which are about $4\%$ and $36\%$ discrepancy between the theoretical and the numerical results. Restricting to $\sigma\leq5$, where the first overtone can be fitted reliably, the relative median difference for $n=1$ drops to $16\%$.   Notice that the first overtone decays about three times faster than the fundamental mode, so it can only be fitted in a short window after the peak, where $n=0$ and the prompt response still dominate the signal and which sources larger innacuracies than for  $n=0$.

The numerical panels also contain a few branches that differ from the theory. For $n=1$, the outer branch along $\sigma$ is only recovered in discontinous segments above $\sigma\simeq6$, at $|\nu|\simeq0.7$. This is next to the region where $|C_1|$ has its largest error. At large $\sigma$ (narrrow band sources in the frequency domain), the excitation of the modes is exponentially suppressed~\cite{Vega:2026lfs}, sourcing a larger uncertainty in our numerical fits. Close to $\nu=0$ and above $\sigma\simeq6.5$, the extraction gives a maximum along $\sigma$ on the inner branch, where the theory has a minimum, and the two inner branches in either side of $\nu=0$ merge with a short minimum near $\sigma\simeq9.5$. The map is almost flat in $\sigma$ there. Other branches are only partly recovered. The minimum of $|C_0|$ along $\nu$ at $\nu=0$ stops at $\sigma\simeq6.5$, since above it $|C_0^{\rm num}|$ is slightly larger than the theory near $\nu=0$. For $n=1$ we recover only pieces of the branch of minima that joins $(0,\sigma_1^*)$ and $(\pm0.28,3.31)$, and the line $\nu=0$ only up to $\sigma\simeq3.5$ as a maximum and up to $\sigma\simeq6.6$ as a minimum.

%%%%%%%%%%%%%%%%%%%%%%%%%%%%%%%%%%%%%%%%%%%%%%%%%%%%%%%%%%
%%%%%%%%%%%%%%%%%%%%%%%%%%%%%%%%%%%%%%%%%%%%%%%%%%%%%%%%%%
\section{Phase landscape: an independent observable}
\label{sec:phase}
%%%%%%%%%%%%%%%%%%%%%%%%%%%%%%%%%%%%%%%%%%%%%%%%%%%%%%%%%%
%%%%%%%%%%%%%%%%%%%%%%%%%%%%%%%%%%%%%%%%%%%%%%%%%%%%%%%%%%

The excitation amplitude characterizes how efficiently a given perturbation excites each QNM, but it does not exhaust the information contained in the complex excitation coefficient $C_n$. Its phase provides an independent probe of the source and the spacetime. We therefore extend the analysis to the phase landscape in the $(\sigma,\nu)$ parameter space and examine how its structure encodes the real and imaginary parts of the QNM frequency.

%%%%%%%%%%%%%%%%%%%%%%%%%%%%%%%%%%%%%%%%%%%%%%%%%%%%%%%%%%
\subsection{Phase decomposition}
\label{sec:phase_structure}
%%%%%%%%%%%%%%%%%%%%%%%%%%%%%%%%%%%%%%%%%%%%%%%%%%%%%%%%%%

Starting from Eq.~\eqref{eq:tn_analytic}, we factor out the common Gaussian envelope 
$e^{-\sigma^2(\omega_n^2+\nu^2)/2}$ to bring the two spectral branches into the 
single hyperbolic factor $\cosh(\sigma^2\omega_n\nu)$,
\begin{equation}
T_n = -i\sqrt{8\pi}\,\sigma\,\omega_n\,
      e^{-\frac{\sigma^2}{2}\big(\omega_n^2 + \nu^2\big)}\, 2\cosh\bigl(\sigma^2\omega_n\nu\bigr).
\label{eq:tn_factored}
\end{equation}
This form separates the phase accumulated by the common spectral envelope from that generated by the coherent combination of the two branches at $\pm\nu$. The phase of $T_n$ then can be decomposed as,
\begin{equation}
\arg\big(T_n\big) = -\frac{\pi}{2}+\theta_\omega+\sigma^2\,\omega_n^{\rm Re}\,\omega_n^{\rm Im}+\Phi_{\rm int}(\sigma,\nu),
\label{eq:arg_Tn}
\end{equation}
where,
\begin{equation}
    \theta_\omega \equiv -\arctan\!\Bigg(\frac{\omega_n^{\rm Im}}{\omega_n^{\rm Re}}\Bigg) = -\arctan\!\Bigg(\frac{1}{\mathfrak{W}_n}\Bigg).
\end{equation}
and the interference phase reads,
\begin{equation}
    \Phi_\mathrm{int}(\sigma,\nu)
    \equiv
    -\arctan\!\Bigg[
        \tanh\!\Big(\sigma^2\omega^{\rm Re}_n\nu\Big)\,
        \tan\!\Big(\sigma^2|\omega^{\rm Im}_n|\nu\Big)
    \Bigg].
    \label{eq:Phi_int}
\end{equation}
For the complete QNEC one has,
\begin{equation}
\phi_n\equiv\arg(C_n)=\arg(B_n)+\arg(T_n)\,.
\label{eq:arg_Cn}
\end{equation}
Since $B_n$ is independent of $(\sigma,\nu)$, it only shifts the absolute phase and does not modify the geometry or stationary structure of the phase landscape.

The \emph{spacetime} contribution $-\pi/2+\theta_\omega$ is independent of the source parameters $(\sigma,\nu)$ Its constant $-\pi/2$ contribution originates from the prefactor $-i$.
For weakly damped modes $\omega_n^{\rm Im} \ll \omega_n^{\rm Re}$ one has $\theta_\omega \to 0$ and this contribution approaches $-\pi/2$, as $-i\omega_n$ becomes purely imaginary.

The \emph{envelope} term $\sigma^2\omega_n^{\rm Re}\omega_n^{\rm Im}$ arises from the imaginary part of $-\sigma^2\omega_n^2/2$ in the common Gaussian envelope shared by both branches. The resulting phase depends on the source width $\sigma$, but not on the carrier frequency $\nu$, and is set by the product $\omega_n^{\rm Re}\omega_n^{\rm Im}$. It survives in the $\nu\to0$ limit for non-oscilatory Gaussian. Thus it corresponds to the phase acquired by the Gaussian source spectrum when evaluated at the complex QNM frequency and vanishes in the limit of a purely real mode frequency. 

The \emph{interference} term $\Phi_{\rm int}(\sigma,\nu)$ contains the entire dependence of the phase on the carrier frequency $\nu$. It arises from the coherent combination of the two spectral branches at $\pm\nu$ and vanishes identically at $\nu=0$. Its structure is controlled by the two dimensionless combinations $\sigma^2\omega_n^{\rm Re}\nu$ and $\sigma^2\omega_n^{\rm Im}\nu$, which enter through the hyperbolic and trigonometric factors in Eq.~\eqref{eq:Phi_int}, respectively. The same two-branch interference responsible for the modulation of the excitation amplitude therefore generates this non-trivial part of the phase. 

%%%%%%%%%%%%%%%%%%%%%%%%%%%%%%%%%%%%%%%%%%%%%%%%%%%%%%%%%%
\subsection{Stationary structure of the phase}
\label{sec:phase_sigma_extrema}
%%%%%%%%%%%%%%%%%%%%%%%%%%%%%%%%%%%%%%%%%%%%%%%%%%%%%%%%%%
The interference phase $\Phi_{\rm int}(\sigma,\nu)$ in Eq.~\eqref{eq:Phi_int} is an even function of $\nu$ that vanishes at $\nu=0$, $\Phi_{\rm int}(\sigma,0)=0$. Expanding around the origin gives,
\begin{equation}
\Phi_{\rm int}(\sigma,\nu)=-\sigma^4\omega_n^{\rm Re}\omega_n^{\rm Im}\,\nu^2+\mathcal{O}(\nu^4),
\end{equation}
and therefore,
\begin{equation}
\partial_{\nu\nu}\arg(T_n)\bigg|_{\nu=0}=-2\sigma^4\omega_n^{\rm Re}\omega_n^{\rm Im}<0.
\label{eq:phase_curvature_nu}
\end{equation}
Hence, $\nu=0$ is a local maximum of the phase along the $\nu$ direction for every $\sigma$.
Defining,
\begin{align}
\mathcal{F}_n \equiv \omega_n^{\rm Im}\sinh\Big(2\sigma^2\omega_n^{\rm Re}\nu\Big)+ \omega_n^{\rm Re}\sin\Big(2\sigma^2\omega_n^{\rm Im}\nu\Big),
\label{eq:Qn_def}
\end{align}
the derivative of the phase with respect to $\nu$ can be written as,
\begin{equation}
\partial_\nu\arg(T_n)=-\sigma^2\frac{\mathcal{F}_n}{\mathcal{D}_n}.
\label{eq:phase_nu_derivative}
\end{equation}
The stationarity condition at fixed $\sigma$ is therefore $\mathcal{F}_n=0$. For $\nu\geq0$,
\begin{align}
\omega_n^{\rm Im}\sinh\!\Big(2\sigma^2\omega_n^{\rm Re}\nu\Big)&\geq2\sigma^2\omega_n^{\rm Re}\omega_n^{\rm Im}\nu,\\[0.5em]
\omega_n^{\rm Re}\sin\!\Big(2\sigma^2\omega_n^{\rm Im}\nu\Big)&\geq-2\sigma^2\omega_n^{\rm Re}\omega_n^{\rm Im}\nu,
\end{align}
where equality in both relations occur only at $\nu=0$. Therefore, $\mathcal{F}_n=0$ at $\nu=0$, while $\mathcal{F}_n >0$ for every $\nu>0$. Together with the $\nu-$antisymmetry of $\mathcal{F}_n$, this implies $\mathcal{F}_n<0$ for $\nu<0$, so that $\nu=0$ is the unique stationary point along the $\nu$ direction. The $\nu<0$ sector is the reflection of the $\nu>0$ sector and contains no additional independent stationary solutions. Unlike the amplitude, whose central extremum can change character as the source width varies, the phase retains the same carrier-frequency extremum for every width and every mode considered.

Varying $\sigma$ changes both the envelope contribution $\sigma^2\omega_n^{\rm Re}\omega_n^{\rm Im}$ and the interference phase. Explicitly,
\begin{equation}
\partial_\sigma\arg(T_n) = 2\sigma\Bigg[ \omega_n^{\rm Re}\omega_n^{\rm Im} - \nu\frac{\mathcal{F}_n}{\mathcal{D}_n} \Bigg].
\label{eq:phase_sigma_derivative}
\end{equation}
The stationarity condition at fixed $\nu$ therefore becomes,
\begin{equation}
\nu\mathcal{F}_n = \omega_n^{\rm Re}\omega_n^{\rm Im}\, \mathcal{D}_n.
\label{eq:phase_sigma_extremum}
\end{equation}
In the large-width limit, the stationary condition with respect to $\sigma$ simplifies because the hyperbolic contributions dominate the bounded trigonometric terms. For fixed $\nu>0$, one has $\mathcal{D}_n\sim \frac{1}{2}\exp(2\sigma^2\omega_n^{\rm Re}\nu)$ and $\mathcal{F}_n\sim \frac{1}{2}\omega_n^{\rm Im}\exp(2\sigma^2\omega_n^{\rm Re}\nu)$, so that $\mathcal{F}_n/\mathcal{D}_n\to\omega_n^{\rm Im}$. Eq.~\eqref{eq:phase_sigma_extremum} therefore yields $\nu\omega_n^{\rm Im}\to\omega_n^{\rm Re}\omega_n^{\rm Im}$, and hence the phase-stationary branch approaches $\nu=\omega_n^{\rm Re}$ as $\sigma\to\infty$. This asymptotic result differs from the amplitude-stationary branches, which approach the linewidth boundaries $\nu=\omega_n^{\rm Re}\pm\omega_n^{\rm Im}$. The phase thus selects the oscillation frequency in the narrowband limit, whereas the amplitude-stationary structure also reflects the damping rate. This difference reflects the complementary information carried by the two observables. The complex QNM spectrum is thus encoded differently in the amplitude and phase landscapes making it truly an independent observable.

\begin{figure*}[!ht]
\centering
\includegraphics[width=\textwidth]{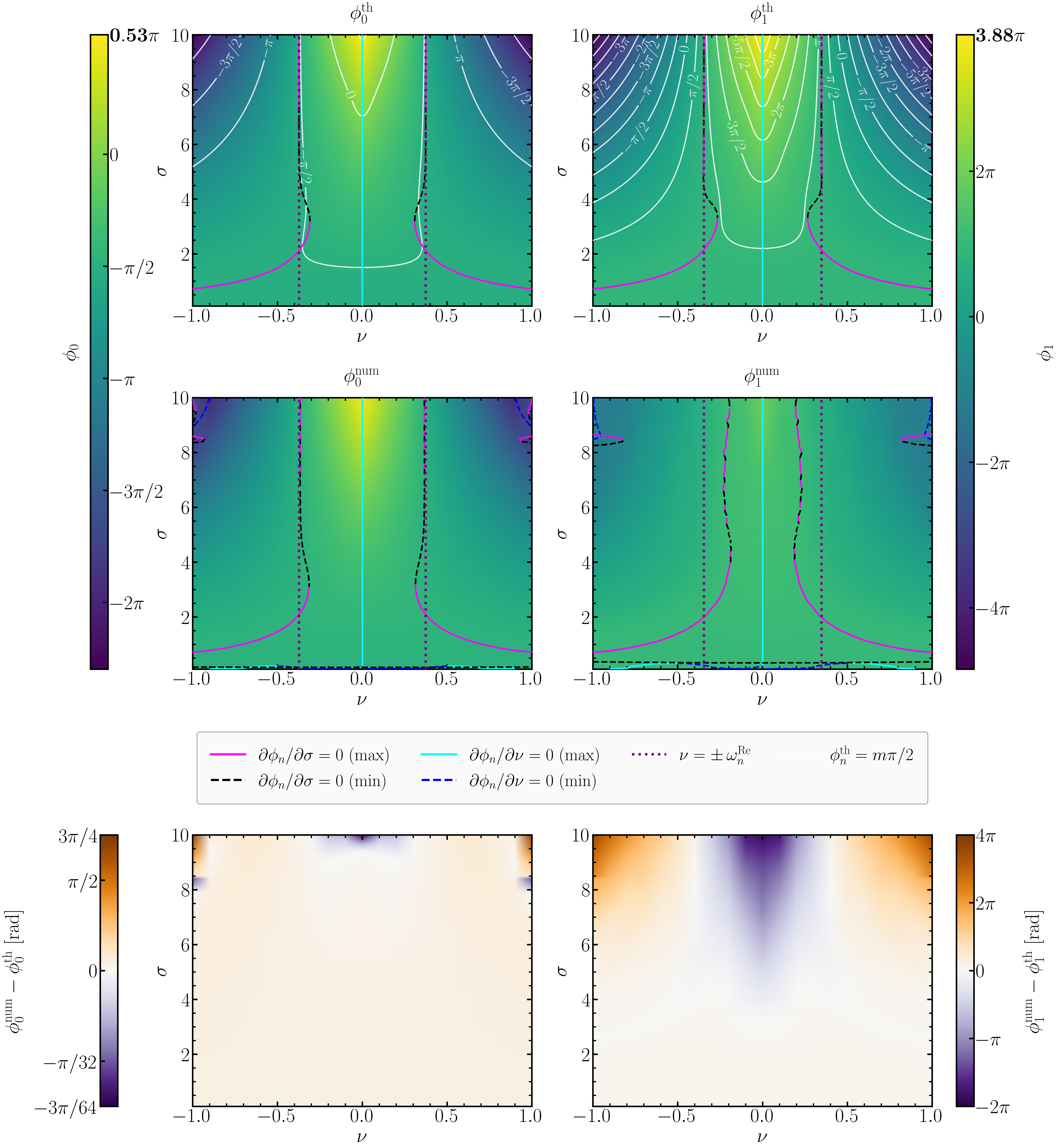}
\caption{\justifying From top to bottom, the rows show the theoretical prediction, the numerical result, and the corresponding phase difference, in radians and with its own colour scale for each mode. In each row, the left panel corresponds to $n=0$ and the right panel to $n=1$. White curves are the theoretical contours $\phi_n=m\pi/2$. Magenta solid and black dashed curves are maxima and minima along $\sigma$, and cyan solid and blue dashed curves are maxima and minima along $\nu$. The purple dotted lines are $\nu=\pm\omega_n^{\rm Re}$.}
\label{fig:unwrapped_phi_heatmaps}
\end{figure*}

%%%%%%%%%%%%%%%%%%%%%%%%%%%%%%%%%%%%%%%%%%%%%%%%%%%%%%%%%%
\subsection{Numerical validation of the phase landscape.}
%%%%%%%%%%%%%%%%%%%%%%%%%%%%%%%%%%%%%%%%%%%%%%%%%%%%%%%%%%

In Fig.~\ref{fig:unwrapped_phi_heatmaps} we show the unwrapped phase landscape $\phi_n=\arg(C_n)$ of Eq.~\eqref{eq:arg_Cn},  in the $(\sigma, \nu)$ plane for the fundamental ($n=0$, left) and the first overtone ($n=1$, right). The top row shows the analytical prediction, while the middle row shows the phase extracted from the numerical time-domain evolutions. The bottom row displays the corresponding numerical-theoretical phase difference. White curves in the theoretical panels denote contours of constant phase for different integer $m$, where $\phi=m\pi/2$. The overlaid colored curves identify the stationary structure of the phase where solid magenta and black dashed curves correspond to maxima and minima with respect to $\sigma$, respectively, while solid cyan and dashed blue curves correspond to maxima and minima with respect to $\nu$ (see Eq.~\eqref{eq:phase_nu_derivative}). In the numerical validation plots (middle row), the extrema are obtained by directly imposing the corresponding stationary conditions and determining their properties from the numerical data. The purple dotted lines mark the QNM frequencies $\nu=\pm\omega_n^{\rm Re}$. A three-dimensional representation of these phase landscapes are also available online at~\cite{SBHR3D}.

The solid cyan branch at $\nu=0$ is common to both modes. As shown in Sec.~\ref{sec:phase_sigma_extrema}, the phase increases monotonically for $\nu<0$ and decreases also monotonically for $\nu>0$, so that $\nu=0$ is the unique maximum along every fixed-$\sigma$ slice. Unlike the maxima of the QNEC amplitudes, it does not move towards $\nu=\pm\omega_n^{\rm Re}$ with increasing narrowband sources. It becomes increasingly pronounced with increasing $\sigma$ and with the mode's damping.
The stationary structure along the $\sigma$ direction is richer. The solid magenta and dashed black branches alternate between maxima and minima as $\sigma$ increases. In the large-$\sigma$ (narrow band) limit, these branches progressively asymptote toward $\nu=\pm\omega_n^{\rm Re}$, marked by the purple dotted lines. 

The two modes show different phase structures away from the stationary branches. For the fundamental mode, the phase varies relatively slowly over most of the parameter space, as reflected by the broader separation between neighbouring constant-phase contours. For the first overtone, these contours become progressively denser as $\sigma$ and $\nu$ increase, indicating a faster phase variation. The characteristic variation scale in the carrier frequency follows $\Delta \nu \propto 1/(\sigma^2 \omega_n^{\rm Im})$. This behaviour originates from the interference contribution and is enchanced by the larger value of $\omega_1^{\rm Im}$.

For $n=0$ the numerical phase follows the theoretical predictions over the whole plane, with a median difference of $0.14\, \rm rad$, which is sustained up to $\sigma\simeq6$. At larger $\sigma$ it stays between $-0.14$ and $+0.25\,{\rm rad}$, except at the edges $|\nu|=1$ for $\sigma\gtrsim8.5$, where it reaches about $+2.2\,{\rm rad}$. Again, the strong suppression of the QNM response makes the signal be dominated by a two orders of magnitude larger prompt, which sources the numerical differences on our numerical results.
For $n=1$ we obtain a good agreement between theoretical and numerical results up to $\sigma\simeq5$, with a median difference of $0.34\,{\rm rad}$, and it degrades quickly above it. At large $\sigma$ the difference is positive towards the upper corners and negative around $\nu\simeq0$, changing the sign across $\nu$. In both regions it goes beyond $\pm\pi$, up to about $3.6\pi$ in the corners and $-1.9\pi$ near $\nu=0$ at $\sigma=10$. Thus, the first overtone shows larger and more structured local discrepancies in the large-$\sigma$ regime, where the theoretical phase varies most rapidly and where $|C_1|$ is also least accurate (see Fig.~\ref{fig:Cn_heatmaps}). 

Finally, some extrema curves in the numerical phase landscapes have no theoretical counterparts. For instance, for both modes, an additional minimum branch in the $\sigma$ space appears along the lower boundary at $\sigma\simeq 0 $ in $\phi_0^{\rm num}$ and $\phi_1^{\rm num}$ as shown in the middle panels in black dashed. Moreover, also in the low $\sigma$ regime, we observe two numerical maxima and minima curves (solid cyan and dashed blue) not supported by our analytical estimates. In this regime, notice that the derivatives with respect to $\nu$ vanish  at $\sigma =0$ (Eq.~\eqref{eq:phase_nu_derivative}) therefore, one may expect that small residual noise displaces slightly the position of the maxima/minima compared to their theoretical counterpart. Other numerical maxima/minima appear in the large $\nu$ large $\sigma$ regime. In this regime, as shown in Fig.~\ref{fig:Cn_heatmaps}, the excitation amplitudes $|C_0|$ and $|C_1|$ is strongly suppressed by: (i) the large mismatch between the carrier frequency $\nu$ and the QNM oscillation frequency $\omega_n^{\rm Re}$ and (ii) the exponential suppression induced by the Gaussian envelope. Together, these effects significantly shorten the time windows available for fitting the QNM contributions, leading to larger numerical errors than in regions with smaller $\sigma$ and $\nu$.  
%%%%%%%%%%%%%%%%%%%%%%%%%%%%%%%%%%%%%%%%%%%%%%%%%%%%%%%%%%
%%%%%%%%%%%%%%%%%%%%%%%%%%%%%%%%%%%%%%%%%%%%%%%%%%%%%%%%%%
\section{Discussion and Conclusion}
\label{sec:conclusions}
%%%%%%%%%%%%%%%%%%%%%%%%%%%%%%%%%%%%%%%%%%%%%%%%%%%%%%%%%%
%%%%%%%%%%%%%%%%%%%%%%%%%%%%%%%%%%%%%%%%%%%%%%%%%%%%%%%%%%
In this work, we have extended the spectral-filtering picture of BH ringdown developed in Paper~I to the full complex QNEC-parameter space. By varying the Gaussian width $\sigma$ and carrier frequency $\nu$, we have mapped how the source spectrum couples to individual QNMs across the $(\sigma,\nu)$ space. Two ingredients organize this response. First, the source spectrum undergoes a bifurcation at $\alpha=\sigma|\nu|=1$, separating a single-peak low-frequency regime from a frequency-selective regime with two resolved branches. Second, these positive- and negative-frequency branches contribute coherently to the QNM overlap, generating interference effects that shape both the excitation amplitude and its phase.

The main results can be summarized as follows. First, we observe that the strength of the BH response $|C_n|$, for our oscillatory Gaussian pulse, is determined by both the spectral width $\sigma$ and the carrier frequency $\nu$ across the $\nu,\sigma$ space (see also Paper~I). In the amplitude landscape, $|C_n|$, we have found that the stationary structure along the carrier-frequency $\nu$ depends on the damping of the mode $\omega_n^{\rm Im}$. For  weakly damped modes ($\mathcal{P}_n>0$, i.e. $n=0,1$ for $\ell=2$), the $\nu=0$ extremum changes character from relative maximum to relative minimum at the critical width $\sigma=\sigma_n^*$, while for $n>1$, i.e, $\mathcal{P}_n<0$, it remains a maximum as shown in Eq.~\eqref{eq:hessian_nu}. In the $\sigma \rightarrow \infty$ limit ($\alpha > 1$), the nonzero-$\nu$ maxima approach $\nu=\pm\omega_n^{\rm Re}$, whereas the stationary branches along the $\sigma$ direction asymptotically approach the  linewidth positions $\omega_n^{\rm Re}\pm\omega_n^{\rm Im}$. Notably, for sufficiently broadband sources, i.e. small $\sigma$ ($\sigma<\sigma_n^*$ for $n=0,1$), the maximum QNM excitation need not occur at $\nu=\pm\omega_n^{\rm Re}$, showing that efficient QNM excitation does not require direct matching between the source carrier frequency and the QNM oscillation frequency as shown in Fig.~\ref{fig:Cn_heatmaps}. Moreover, we observe that the interference between the positive- and negative-frequency branches of the source introduces an oscillatory modulation of the excitation amplitude, with a characteristic period set by the QNM damping rate. This modulation is negligible for the  $n=0$ but becomes increasingly pronounced for $n\geq 1 $, producing additional local minima and a richer stationary structure in the excitation landscape of $|C_{n=2}|$ and $|C_{n=3}|$. 

In the phase landscape, $\nu=0$ remains the unique maximum for fixed $\sigma$, while the stationary branches in $\sigma$ asymptotically reaches at $\nu=\pm\omega_n^{\rm Re}$ in the narrowband limit. The amplitude and phase therefore encode complementary information about the real and imaginary parts of the QNM frequency. These results complement investigations of QNM excitation that have focused on the intrinsic excitation factors $B_n$~\cite{Berti:2006wq, Zhang:2013ksa,Oshita:2021iyn,Lo:2025njp,DellaRocca:2025zbe,Glampedakis:2003dn,Chen:2024hum} and their resonant enhancement~\cite{Motohashi:2024fwt,Takahashi:2025uwo,Kubota:2025hjk}, and also beyond Kerr~\cite{Rossi:2026als,Zhao:2026nce}. Furthermore, the interference between the positive- and negative-frequency branches of the perturbation reveals a novel mechanism shaping QNM excitation, complementary to the destructive interference among distinct QNMs investigated in the context of spectral instabilities~\cite{Oshita:2025ibu}. They also build on earlier work on how the source sets the QNM amplitudes, from ID~\cite{Leaver:1986gd,SunPrice1988,Andersson:1995zk,Andersson:1996cm,Nollert:1998ys,Dorband:2006gg,Dolan:2011fh,Oshita:2024wgt,Chavda:2024awq} to infalling particles~\cite{davis:1971gg,Mino:2008at,Price:2015gia,Hughes:2019zmt}, and on the QNM expansion that defines the QNECs~\cite{Nollert:1992ifk,Casals:2013mpa,Ansorg:2016ztf,Green:2022htq,Besson:2024adi}.

The analytical structure derived throughout this work is supported by the time-domain RW evolutions. The extraction reproduces the main stationary features of both the amplitude and phase landscapes over most of the $(\sigma,\nu)$ parameter space, with the largest discrepancies appearing in the rapidly varying large-$\sigma$ regime, particularly for the first overtone. For $n=0$ the median difference is about $4\%$ in amplitude and $0.14\,{\rm rad}$ in phase. For $n=1$ it is about $16\%$ and $0.34\,{\rm rad}$ up to $\sigma\simeq5$, and $36\%$ in amplitude over the whole plane. Notice that at larger widths, the stronger suppression of the overtone makes its numerical extraction increasingly susceptible to errors. Finally, for the $n>1$ modes with short damping times, we haven't been able to obtain stable numerical fits.  Overall, the results show that the QNECs in the source-parameter space can be viewed as a spectroscopic map of the QNM excitation. The properties of the perturbing source and the intrinsic complex QNMs are encoded together in the QNECs. This provides a useful framework for studying how different source spectra probe BH resonances. An important next step is to extend this analysis to more general perturbations and physically motivated sources. Near-circular plunges close to the innermost stable circular and eccentric orbits excite the ringdown resonantly~\cite{Ori:2000zn, Hadar:2009ip, Apte:2019txp, Lim:2019xrb, Taracchini:2014zpa, Nasipak:2019hxh, Thornburg:2019ukt,Carullo:2024smg,DellaRocca:2025zbe,DeAmicis:2025xuh}, with direct relevance for extreme mass-ratio inspirals~\cite{amaroseoane:2007aw, 2017arxiv170200786a}. These aspects will be explored in the future.

%%%%%%%%%%%%%%%%%%%%%%%%%%%%%%%%%%%%%%%%%%%%%%%%%%%%%%%%%%
%%%%%%%%%%%%%%%%%%%%%%%%%%%%%%%%%%%%%%%%%%%%%%%%%%%%%%%%%%
\section*{Acknowledgements}
%%%%%%%%%%%%%%%%%%%%%%%%%%%%%%%%%%%%%%%%%%%%%%%%%%%%%%%%%%
%%%%%%%%%%%%%%%%%%%%%%%%%%%%%%%%%%%%%%%%%%%%%%%%%%%%%%%%%%

This work was supported by the Universitat de les Illes Balears (UIB) with funds from the Programa de Foment de la Recerca i la Innovaci\'o de la UIB 2024-2026 (supported by the yearly plan of the Tourist Stay Tax ITS2023-086); the Spanish Agencia Estatal de Investigaci\'on grants PID2022-138626NB-I00, RED2024-153978-E, RED2024-153735-E, funded by MICIU/AEI/10.13039/501100011033 and the ERDF/EU; and the Comunitat Aut\`onoma de les Illes Balears through the Conselleria d'Educaci\'o i Universitats with funds from the ERDF (SINCO2022/18146 - Plataforma HiTech-IAC3-BIO). Project funded by the Comunitat Autonòma de les Illes Balears through a "Ajut per a projectes de recerca científica i tecnològica (PRD2025, PRD$2025\_099$)".
X.J.\ is supported by the Spanish Ministerio de Ciencia, Innovaci\'on y Universidades (Beatriz Galindo, BG22-00034) and cofinanced by UIB.
S.D.\ acknowledges financial support from MUR, PNRR - Missione~4 - Componente~2 - Investimento~1.2 - finanziato dall'Unione europea - NextGenerationEU (cod.\ id.:\ SOE2024\_0000167, CUP: D13C25000660001).
A.S.K.\ gratefully acknowledges the Gran Sasso Science Institute (GSSI) for its hospitality during the research stay and Andrea Maselli for his academical support.
A.S.K.\ also acknowledges support from the Conselleria d'Educaci\'o i Universitats of the Govern de les Illes Balears through the ``Ajuts a la mobilitat del personal investigador de les Illes Balears 2025--2028'' (MOB\_036\_26), within project ITS 163-2019 ``Formaci\'o i Incorporaci\'o de Doctors al Sistema de Ci\`encia i Tecnologia de les Illes Balears'', funded through the Impost de Turisme Sostenible.
The authors thankfully acknowledge the computer resources at MareNostrum~5 and the technical support provided by the Barcelona Supercomputing Center (BSC) through grants No.~RES-AECT-2025-1-0011, RES-AECT-2025-2-0038, RES-AECT-2025-3-0050, and RES-AECT-2026-1-0063 from the Red Espa\~nola de Supercomputaci\'on (RES).
The authors acknowledge CINECA for providing high-performance computing resources and support through the ISCRA initiative under project HP10CU7X29.

%\clearpage

\bibliography{main}

\end{document}